\documentclass[a4paper,12pt]{article}
\usepackage{jheppub,booktabs}
\usepackage{amsmath,amssymb,slashed}
\usepackage{braket}
\usepackage{xcolor,colortbl}
\usepackage{cancel}
\usepackage{subfigure}
\usepackage{mathrsfs}
\usepackage{graphicx}
\usepackage[compat=1.1.0]{tikz-feynman}
\usepackage{pdflscape}
\usepackage{hepunits}
\usepackage{feynmf}
\usepackage{enumerate}
\usepackage[utf8x]{inputenc}
\usepackage{xspace}
\usepackage{hyperref}

\newcommand{\SMEFT}{\textsc{SMeft}\xspace}
\newcommand{\BSMEFT}{\textsc{IDMeft}\xspace}

\newcommand{\dd}{\mathrm{d}}
\newcommand{\defEq}{\overset{\mathrm{def}}{=}}

\newcommand{\MeV}{\, \mathrm{MeV}}
\newcommand{\GeV}{\, \mathrm{GeV}}
\newcommand{\TeV}{\, \mathrm{TeV}}
\newcommand{\PeV}{\, \mathrm{PeV}}

\newcommand{\cm}{\, \mathrm{cm}}

\newcommand{\dist}{\hspace*{4mm} , \hspace*{4mm}}

\newcommand{\calchep}{\texttt{CalcHEP}\xspace}
\newcommand{\micromegas}{\texttt{micrOMEGAs}\xspace}
\newcommand{\DsixTools}{\texttt{DsixTools~2.0}\xspace}
\newcommand{\lambdaOne}{\lambda_{1}}
\newcommand{\lambdaTwo}{\lambda_{2}}
\newcommand{\lambdaThree}{\lambda_{3}}
\newcommand{\lambdaFour}{\lambda_{4}}
\newcommand{\lambdaFive}{\lambda_{5}}

\newcommand{\lambdaTFF}{\lambda_{345}}

\newcommand{\muTwo}{\mu_{2}}

\newcommand{\Omegahh}{\Omega h^{2}}

\author[b,a]{María Dias Astros,}
\author[a]{Sven Fabian,}
\author[a]{and Florian Goertz}
\affiliation[a]{Max-Planck-Institut f{\"u}r Kernphysik\\ Saupfercheckweg 1, 69117 Heidelberg, Germany}
\affiliation[b]{Physikalisches Institut, Albert-Ludwigs-Universit{\"a}t Freiburg\\ Hermann-Herder-Straße 3, 79014 Freiburg, Germany}
\emailAdd{maria.dias@physik.uni-freiburg.de}
\emailAdd{sven.fabian@mpi-hd.mpg.de}
\emailAdd{florian.goertz@mpi-hd.mpg.de}

\title{Minimal Inert Doublet Benchmark\\ for Dark Matter and the Baryon Asymmetry
}

\abstract{\\ 
In this article we discuss a minimal extension of the Inert Doublet Model (IDM) with an effective $CP$-violating $D=6$ operator, involving the inert Higgs and weak gauge bosons, that can lift it to a fully realistic setup for creating the  baryon asymmetry of the Universe (BAU). Avoiding the need to stick to an explicit completion, we investigate the potential of such an  operator to give rise to the measured BAU during a multi-step electroweak phase transition (EWPhT) while sustaining a viable DM candidate in agreement with the measured relic abundance. We find that the explored extension of the IDM can account quantitatively for both DM and for baryogenesis and has quite unique virtues, as we will argue. It can thus serve as a benchmark for a minimal realistic extension of the SM that solves some of its shortcomings and could represent the low energy limit of a larger set of viable completions.

After discussing the impact of a further class of operators that open the possibility for a larger mass splitting (enhancing the EWPhT) while generating the full relic abundance also for heavy inert-Higgs DM, we ultimately provide a quantitative evaluation of the induced lepton electric dipole moments in the minimal benchmark for the BAU. These arise here at the two-loop level and are therefore less problematic compared to the ones that emerge when inducing $CP$ violation via an operator involving the SM-like Higgs.}
\begin{document}
\maketitle

\section{Introduction and model setup} \label{Sec:Intro}
Thanks to the discovery of a resonance resembling the Higgs particle proposed in the 1960s~\cite{Higgs_1964,Englert_1964,Guralnik_1964}, at ATLAS~\cite{Aad_2012} and CMS~\cite{Chatrchyan_2012} at the Large Hadron Collider in 2012, the minimal Standard Model of Particle Physics (SM) was completed. Subsequent studies of couplings of the Higgs particle to fermions and electroweak (EW) gauge bosons showed agreement with the SM predictions and thus demonstrated, once more, the powerful predictiveness of the theory. However, despite the success of gaining understanding of the properties of elementary particles and their interactions, it is well-known that the SM lacks in providing explanations for various phenomena, \emph{inter alia}, the existence of dark matter (DM)  and the observed baryon asymmetry of the Universe (BAU). 

In this article, we attempt to address these questions via an effective-field-theory (EFT) approach for the Inert Doublet Model (IDM), minimally extended at a beyond-IDM energy scale~$\Lambda$. The IDM has already been widely studied as a model for DM and in the context of the electroweak phase transition (EWPhT) as a first step towards explaining baryogenesis~\cite{Ginzburg_2010,Gil_2012,Blinov_2015,Blinov_2015_2step,Fabian_2021}. Nonetheless, since the interactions between the additional inert scalars and the SM states preserve the $CP$ symmetry, baryogenesis cannot be achieved in the non-modified `vanilla' IDM. Adding an effective $CP$-violating operator allow us then to (quantitatively) accommodate the missing Sakharov condition and explain the BAU within the framework. 
Moreover, due to its minimal nature, this effective IDM could serve as a realistic economic benchmark extension of the SM that solves prominent shortcomings and -- with its new scalars being preferably rather light -- could be seen as the low energy limit of a larger class of viable completions residing at higher scales.
 
The existence of DM is well established through a wide range of observations~\cite{Markevitch_2004,Rubin_1980,Planck_2020}, including colliding clusters (\emph{e.g.} the bullet cluster), rotational curves of various galaxies, gravitational lensing, structure formation, big bang nucleosynthesis, and the cosmic microwave background. The energy density of the unknown DM component today is quantified by~\cite{Zyla_2020}
\begin{align}
    \Omegahh_{\mathrm{ref}} \defEq \frac{\rho_{\mathrm{DM,ref}}}{\rho_{\mathrm{crit}}}h^{2} = 0.1200(12) \label{Eq:RelicAbundanceLiterature}
\end{align}
with the critical energy density $\rho_{\mathrm{crit}} = 3H_{0}^{2}/(8\pi G_{N})$ defined in terms of today's Hubble parameter $H_{0}$ and Newton's gravitational constant $G_{N}$. To avoid an overclosure of the Universe, the model under consideration should not predict a larger DM relic abundance than the reference value given above.

Many candidates have been proposed to account for DM\footnote{The DM candidate must be $(i)$ electrically neutral or milli-charged at the most, $(ii)$ at most weakly interacting with SM particles and
$(iii)$ stable on cosmological time scales. 
}, among which the weakly interacting massive particles (WIMPs) are of the most appealing. Their mass ranges between a few GeV and $\mathcal{O}(100) \TeV$ and they interact only weakly with the SM particles. WIMPs are thermally produced via freeze out
and their ``final" comoving density can make out the entirety of the measurable DM relic abundance in Eq.~(\ref{Eq:RelicAbundanceLiterature}).
The IDM naturally features a WIMP DM candidate. 

Concretely, the IDM (see, \emph{e.g.}, Refs.~\cite{Deshpande_1978,Honorez_2007,Ginzburg_2010,Chowdhury_2012,Borah_2012,Gil_2012,Gustafsson_2012,Blinov_2015,Blinov_2015_2step,Ilnicka_2016,Kalinowski_2018,Fabian_2021,Kalinowski_2021}) is an extension of the SM with an additional $SU(2)_{L}$ doublet scalar~$H_{2}$, odd under a new $\mathbb{Z}_{2}$ symmetry and featuring a vanishing vacuum expectation value (vev) at zero temperature (which guarantees its inert nature, see below). The $SU(2)_{L}$ doublets read 
\begin{align}
    H_{1} = \frac{1}{\sqrt{2}} \begin{pmatrix} \sqrt{2} G^{+} \\
    v_{1} + h + iG \end{pmatrix} \dist H_{2} = \frac{1}{\sqrt{2}} \begin{pmatrix} \sqrt{2} H^{+} \\
    v_{2} + H + iA \end{pmatrix}\,,
    \label{eq:Doublet-expansion}
\end{align}
with the SM Higgs boson $h$ and the vevs $v_{1} \equiv v = 246\GeV$ and $v_{2}=0$ at zero temperature. Both doublets are $\left(\textbf{2},1\right)$ representations of the EW $SU(2)_{L}\times U(1)_{Y}$ gauge group and the Goldstone bosons $G$, $G^{\pm}$ are associated with the longitudinal modes of the respective EW gauge bosons $Z$, $W^{\pm}$ after EW symmetry breaking. Similarly, $H^{\pm}$ correspond to two new $CP$-even, electrically charged physical scalars, whereas $H$ and $A$ are two additional neutral scalars, the former being $CP$-even while the latter is $CP$-odd. We choose $H$ to be the lightest scalar and therefore the DM candidate. Its stability is guaranteed by the aforementioned $\mathbb{Z}_{2}$ symmetry, under which all SM fields are even but $H_2$ is odd. This is the prominent feature of the IDM and prohibits any interaction term between the inert doublet $H_2$ and SM fermions (and therefore perilous contributions to flavour-changing neutral currents~\cite{Honorez_2007}) at the renormalizable level. 

The scalar potential is given by
\begin{align}
    V(H_1,H_2)  = \  &\mu_{1}^{2} \left\vert H_{1} \right\vert^{2} + \mu_{2}^{2} \left\vert H_{2} \right\vert^{2} + \lambda_{1} \left\vert H_{1} \right\vert^{4} + \lambdaTwo \left\vert H_{2} \right\vert^{4}  \nonumber \\ &+ \lambdaThree \left\vert H_{1} \right\vert^{2}\left\vert H_{2} \right\vert^{2} 
  + \lambdaFour \left\vert H_{1}^{\dagger}H_{2} \right\vert^{2}+ \frac{1}{2} \left[ \lambdaFive \left( H_{1}^{\dagger}H_{2} \right)^{2}+ \mathrm{h.c.}\right]
\end{align}
where all the couplings are real\footnote{The coupling parameter $\lambdaFive$ can be complex in general. However, its complex phase can be removed by a suitable (global) Higgs doublet redefinition.} and the masses of the scalars are given by 
\begin{align}
    m_h^2 &= 2 \lambdaOne v^{2} \ , \ m_H^2 = \muTwo^2 + \lambdaTFF\frac{v^{2}}{2} \ , \ m_A^2 = \muTwo^2 + \bar{\lambda}_{345}\frac{v^{2}}{2} \ , \
    m_{H^{\pm}}^2 =\mu_2^2 +\lambda_3 \frac{v^{2}}{2}  \label{Eq:IDMParameterRelation}
\end{align} 
with the short-hand notations
\begin{equation}
    \lambdaTFF = \lambdaThree + \lambdaFour + \lambdaFive \quad , \quad \bar{\lambda}_{345} = \lambdaThree + \lambdaFour - \lambdaFive \ .
    \label{eq:lam345}
\end{equation} 
The theoretical and experimental constraints on the model, \emph{e.g.} from perturbative unitarity, vacuum stability, invisible SM Higgs decays into a pair of inert scalars, or electroweak precision tests, as well as the parameter space allowing for the correct DM abundance can be found in Ref.~\cite{Fabian_2021} and references therein.

Since real couplings prevent $CP$ violation, the IDM must be augmented in order to become a realistic model of baryogenesis. To this end, we focus mostly on the dimension-six operator
\begin{equation}
    \mathcal{L}_{\mathrm{CPv}}^{\mathrm{IDM}} \supset \mathcal{L}_{\mathrm{BAU}}=\tilde{c}_2 |H_2|^2 V_{\mu\nu}\widetilde{V}^{\mu\nu} \equiv \frac{\tilde{c}_2}{2} \epsilon^{\mu\nu\alpha\beta} \left\vert H_{2} \right\vert^{2} V_{\mu\nu} V_{\alpha\beta} \ , 
    \label{eq:dual-dualOperator}
\end{equation}
which plays a rather unique role within the set of potential operators, as we will explain further below.\footnote{See Refs.~\cite{Grzadkowski:2010au,Krawczyk:2015xhl,Cordero-Cid:2020yba} for different explicit extensions of the IDM with new states to implement $CP$ violation.} Here, $V_{\mu\nu}\widetilde{V}^{\mu\nu}=W_{\mu\nu}^{a}\widetilde{W}^{a,\mu\nu} + B_{\mu \nu}\widetilde{B}^{\mu\nu}$ represents the sum of products of the $SU(2)_{L}$ isospin and $U(1)_{Y}$ hypercharge field strength tensors and their respective duals. Lifting the assumption of equal coefficients does not change the result for the BAU, as this is governed only by the coefficient of the $SU(2)_{L}$ term. As we will see, considering the field strength coupling to the inert doublet has phenomenological advantages over the alternative involving the `active' $H_{1}$ doublet, for example leading to suppressed electric dipole moments of leptons ($\ell$EDMs). Still, also the latter operator can lead to viable results for the BAU in corners of the parameter space, and we will analyze this below, too. However, the focus is on the role of the operator in Eq.~(\ref{eq:dual-dualOperator}) in baryogenesis and its impact on the DM relic abundance, which will be studied in detail in the next sections. 

We point out that the IDM augmented by this operator delivers a minimal, yet versatile, benchmark model to study the simultaneous realization of DM and the BAU. Given stringent constraints on $CP$-violating operators from limits on $\ell$EDMs, together with the modest required corresponding coefficients to realize the BAU that we find, it is very reasonable that $CP$ violation is generated at a higher scale.\footnote{See Refs.~\cite{Baldes:2018nel,Meade:2018saz,Glioti:2018roy,Matsedonskyi:2020mlz} for scenarios where also the EWPhT is lifted to such higher scales.} This makes the implementation via effective operators particularly suitable, being able to describe the effect of a set of potential completions.

The structure of this article is as follows. After having introduced and motivated the model in Sec.~\ref{Sec:Intro}, the baryogenesis mechanism is elaborated on in Sec.~\ref{Sec:EWBG}, including also a discussion on constraints from experimental limits on $\ell$EDMs. Consecutively, the dark matter abundance for suitable parameters is investigated in Sec.~\ref{Sec:DM}. We finally present our conclusions in Sec.~\ref{sec:Conc}, while a series of appendices contains technical details, in particular a two-loop analysis of the $\ell$EDM induced by the second Higgs.

\section{Baryogenesis at the electroweak scale} \label{Sec:EWBG}
In this work we consider the scenario of baryogenesis during the EWPhT, an idea that has been extensively studied in the past (see, \emph{e.g.}, Refs.~\cite{Kuzmin_1985,Shaposhnikov:1986jp,Shaposhnikov:1987tw,Turok_1991,Nelson_1992,Cohen_1993,Quiros_1999,Trodden_1999,Cline_2011,Morrissey_2012,Gan_2017,deVries:2017ncy} and references therein). 
The BAU is quantified by 
\begin{align}
    \eta_{\mathrm{ref}} \defEq \frac{n_{B}}{n_{\gamma}} = \left( 6.143 \pm 0.190 \right)\cdot 10^{-10} \ , \label{Eq:BaryonAsymmetryLiterature}
\end{align}where $n_{B}$ is the difference of baryon and anti-baryon number densities and $n_{\gamma}$ is the number density of photons.
Assuming $CPT$ invariance, the three vital ingredients for successful baryogenesis, elaborated by Sakharov~\cite{Sakharov_1991} in 1967, must be fulfilled: in addition to violation of baryon number $B$ and of charge conjugation symmetry $C$ as well as of the combination $CP$ of charge-conjugation and parity symmetry, the presence of an out-of-equilibrium process is a requisite.

The first condition of this list is met by the fact that neither baryon nor lepton number is conserved in the SM because of the $U(1)_{B+L}$ anomaly, as shown by 't~Hooft in 1976~\cite{_t_Hooft_1976}. This violation is mediated by sphaleron processes which become effective at sufficiently high temperatures as later realized by Kuzmin, Rubakov and Shaposhnikov~\cite{Kuzmin_1985}. 
In fact, for temperatures below $\sim 10^{13} \GeV$~\cite{Cline_2006}, sphalerons are expected to be active and in thermal equilibrium, effectively preventing the creation of a net baryon number. Below the EW scale, on the other hand, sphaleron processes are Boltzmann suppressed. However, a baryon asymmetry can be created \emph{during} the EWPhT and it will then remain, provided that the sphalerons are quickly turned off thereafter. This is the so-called wash-out condition which requires of a strong first-order
phase transition, generating the out-of equilibrium situation mentioned above, that we know is not provided by the SM since the Higgs mass of $m_h\approx 125\,{\rm GeV}$ is too large. On top of that, the $CP$ violation in the weak sector of the SM is too small to explain the measured BAU even if the other two Sakharov conditions were fulfilled (see Ref.~\cite{Cline_2006} for instance). 
Hence, an SM extension must feature an additional source of $CP$ violation and
a strong first-order EWPhT. 
\begin{figure}[b!]
    \centering
            \begin{tikzpicture}
            \draw[thick,-latex] (0,0) -- (6,0) node[anchor=north west] {$\sqrt{2}\braket{H_{1}}$};
            \draw[thick,-latex] (0,0) -- (0,4.1) node[anchor=south east] {$\sqrt{2}\braket{H_{2}}$};

            \draw[red,line width=0.75mm,-{latex[length=3mm, width=2mm]}] (0,0) -- (5.0,0);
            \fill[red] (5.2,0) circle (1.5mm);

            \draw[blue,line width=0.75mm,-{latex[length=3mm, width=2mm]}] (0,0) -- (0,3.1);
            \draw[blue,line width=0.75mm,-{latex[length=3mm, width=2mm]}] plot [smooth, tension=0.8] coordinates {(0.2,3.3) (3.5,2.6) (5.15,0.18)};
            \fill[blue] (0,3.3) circle (1.5mm);

            \draw[blue,dashed,line width=0.75mm,-{latex[length=3mm, width=2mm]}] (0,0) -- (1.82,1.28);
            \draw[blue,dashed,line width=0.75mm,-{latex[length=3mm, width=2mm]}] (2.4,1.28) -- (5,0.1);
            \fill[blue] (2.1,1.4) circle (1.5mm);

            \node at (-0.6,-0.4) {$\left( 0, 0 \right)$};
            \node at (5.2,-0.5) {$\left( v_{1}, 0 \right)$};
            \node at (-0.8,3.3) {$\left( 0, v_{2} \right)$};
            \node at (2.1,1.9) {$\left( v_{1}^{\prime}, v_{2}^{\prime} \right)$};
            \node[text=red] at (2.5,-0.25) {one-step};
            \node[text=blue,rotate=-12] at (2.1,3.34) {two-step};
            \fill[black] (0,0) circle (1.5mm);
        \end{tikzpicture}
    \caption{Illustration of possible scenarios for the evolution of the EW vacuum. During a two-step EWPhT, a non-trivial intermediate field configuration is possible in principle (dashed line), as found in Ref.~\cite{Benincasa_2022}. As a characteristic feature of the IDM, $\braket{H_{2}}$ disappears for $T\rightarrow 0$ in each case.}
    \label{Fig:EWPhT_overview}
\end{figure}
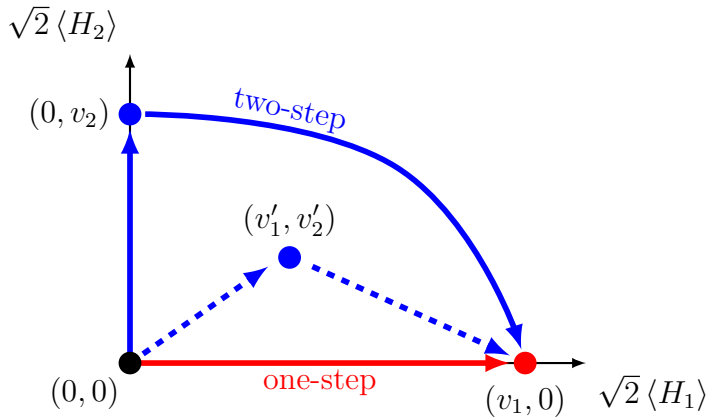

 The latter issue is addressed in the model considered here by the presence of the additional Higgs doublet $H_{2}$, which also opens up the possibility for a multi-step EWPhT (see Fig.~\ref{Fig:EWPhT_overview}). Starting for example in the symmetric phase with vanishing vevs, the scalar potential can evolve either via one transition after which the SM Higgs doublet has developed a finite vev, \emph{i.e.}, $\braket{H_{1},H_{2}} = \left( 0,0 \right) \rightarrow \left( v_{1}/\sqrt{2},0 \right)$, or via (multiple) intermediate steps. Note that the inert nature of $H_{2}$ is restored at zero temperature. We will consider a two-step EWPhT with one additional transition, proceeding as $\braket{H_{1},H_{2}} = \left( 0,0 \right) \rightarrow \left( 0,v_{2}/\sqrt{2} \right) \rightarrow \left( v_{1}/\sqrt{2},0 \right)$, as analyzed recently by two of the present authors in Ref.~\cite{Fabian_2021}.\footnote{Although Ref.~\cite{Benincasa_2022} found recently the possibility of two simultaneous vevs in a slice of parameter space, which would be interesting to examine further, we consider the case of only one finite vev at a time.}

To study the generation of the BAU, first we realize, following the analysis pioneered in Ref.~\cite{Dine_1991}, that the operator introduced in~\eqref{eq:dual-dualOperator} can be written as (see also Refs.~\cite{Cohen_1987,Cohen_1988,Cohen_1991})
\begin{align}
    \tilde{c}_2 \left\vert H_{2} \right\vert^{2} W_{\mu\nu}^{a}\widetilde{W}^{a,\mu\nu} =\frac{32 \pi^2}{3g^2} \tilde{c}_{2} \ j_{B}^{\mu} \ \partial_{\mu} \left\vert H_{2} \right\vert^{2}\,,
    \label{Eq:SBGInteractionTerm}
\end{align}
with the (dual) $SU(2)_{L}$ field strength tensor $W_{\mu\nu}^{a}$ $(\widetilde{W}_{\mu\nu}^{a})$, the beyond-IDM energy scale $\Lambda$ absorbed in the Wilson coefficient $\tilde{c}_{2}\equiv \lambda_{CP}/\Lambda^{2}$, and the baryon current~$j_{B}^{\mu}$.\footnote{We note that, while related operators with scalar multiplets had been considered before (see, \emph{e.g.}, Refs.~\cite{Dine_1991,Dine:1991ck,Gil_2012,Fabian_2021}), the impact of this particular operator on the BAU has not been scrutinized so far.} The interaction term in Eq.~(\ref{Eq:SBGInteractionTerm}) leads to an \emph{effective} chemical potential, producing a shift in the energy levels of baryons with respect to antibaryons in the thermal distribution, and the sphaleron processes generate a BAU during a moderate temporal change of $\braket{H_2}$, like during a two-step EWPhT described above. 

The mentioned shift in the free energy leads to a minimum associated to an equilibrium value for the baryon number density of \cite{Dine_1991} 
\begin{equation}
    n_{B}^{\mathrm{eq}} = \tilde c_2 \frac{8 \pi^2}{3 g^2} \partial_0 |H_2|^2 T^2\,.
\end{equation}
The evolution of baryon number then follows a Boltzmann-like equation of the form
\begin{align}
    \frac{\dd n_{B}}{\dd t} = -18\frac{\Gamma}{T^{3}} \left( n_{B} - n_{B}^{\mathrm{eq}} \right), 
\end{align}
with the sphaleron rate given in terms of the weak coupling $\alpha_W\equiv g^2/(4\pi)$ as 
\begin{align}
   \frac{ \Gamma}{V} \sim \mathcal{O}\left(0.1 - 1 \right) \left(\alpha_{W}T \right)^{4} \label{Eq:RateSphaleronHighT} \,.
\end{align}

\begin{figure}[b!]
    \centering
    \includegraphics[width=0.7\textwidth]{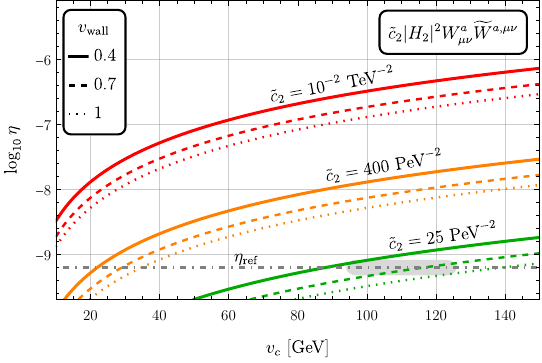}
    \caption{Dependence of the baryon asymmetry parameter $\eta$ on the critical vev $v_{c} = \sqrt{2}\braket{H_2(T_c)}$, bubble wall velocity $v_{\mathrm{wall}}$, and Wilson coefficient $\tilde{c}_{2}$. The bubble wall velocity and BSM coupling are indicated by line style and colour, respectively. The light-gray contour represents the range of possible critical vevs at the first stage of a two-step EWPhT found in Ref.~\cite{Fabian_2021}.}
    \label{Fig:EWBG_H2_asymmetryVSvev}
\end{figure}

Following the estimate in Ref.~\cite{Dine_1991} and considering a strong first-order EWPhT, the resultant BAU in terms of the vev $\langle H_2 \rangle = v_c$ at the critical temperature $T_{c}$ reads 
\begin{align}
     n_B \sim \frac{8 \pi}{3} \tilde{c}_{2} \left\vert v_{c}\right\vert^{2} \alpha_{W}^{4} \ \Delta t \ T_{c}^{4} \ ,
     \label{eq:BaryonNumberDensity}
\end{align}
where $\Delta t$ is the period of time needed by the transition to take place in a volume with a radius given by the correlation length $\xi \sim \left( \alpha_{W} T_{c} \right)^{-1}$. The bubble expansion is assumed to occur with constant velocity $v_{\mathrm{wall}}$, so that $\Delta t = \xi / v_{\mathrm{wall}}$. To quantify the BAU via Eq.~(\ref{Eq:BaryonAsymmetryLiterature}), we recall that the photon number density is given by
\begin{align}
    n_{\gamma} = \frac{\zeta \left( 3 \right)}{\pi^{2}}g_{*}T_{c}^{3} \ ,
    \label{eq:entropy_density}
\end{align}
with the Riemann $\zeta$-function and $g_{*}=2$ spin polarizations, respectively~\cite{Cline_2006}. The resultant dependence of the BAU on the critical vev, the bubble wall velocity\footnote{Note that an ultra-relativistic bubble wall velocity changes the dynamics of the expansion, as studied in Refs.~\cite{Azatov_2021,Azatov_2021_2}.} and the Wilson coefficient $\tilde{c}_{2}$ is shown in Fig.~\ref{Fig:EWBG_H2_asymmetryVSvev}. 
Assuming the new coupling constant to be $\sim \mathcal{O}\left( 1 \right)$, we can inspect that $\Lambda \approx 200\TeV$ results in the measured value of the BAU for a viable value of $v_{c}$ (gray band) and a wide range of bubble wall velocities.

\begin{figure}[b!]
    \centering
    \includegraphics[width=0.7\textwidth]{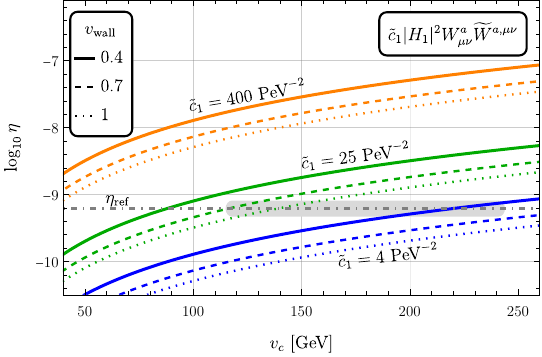}
    \caption{Dependence of the baryon asymmetry parameter $\eta$ on the critical vev $v_{c}=\sqrt{2}\braket{H_1(T_c)}$, bubble wall velocity $v_{\mathrm{wall}}$, and Wilson coefficient $\tilde{c}_{1}$. Like in Fig.~\ref{Fig:EWBG_H2_asymmetryVSvev}, the bubble wall velocity and BSM coupling are indicated by line style and colour, respectively. The light-gray contour represents the range of possible critical vevs of an EWPhT found in Ref.~\cite{Fabian_2021}.} 
    \label{Fig:EWBG_H1_asymmetryVSvev}
\end{figure}

We note that the crucial ingredients of our setup, strong two-step EWPhT and sufficient $CP$ violation, offer promising handles to further probe the framework.
On the one hand, the sizable vevs of the intermediate transition might cause very characteristic gravitational waves signatures (see, \emph{e.g.}, Refs.~\cite{Morais:2019fnm,Liu:2023sey}) whose study is, nonetheless, out of the scope of this work. On the other hand, additional sources of $CP$ violation are in general constrained by null results in measurements of the electric dipole moment of elementary or composite particles like leptons ($\ell$EDM) or baryons (see, \emph{e.g.}, Refs.~\cite{Pospelov_2005,Jungmann_2013,Engel_2013,Jung_2014,Safronova_2018,Cirigliano_2019,Panico_2019,Altmannshofer_2020,Kley_2021}). To ensure that the contribution of the operator defined in Eq.~(\ref{eq:dual-dualOperator}) to the $\ell$EDM is below the sensitivity of ongoing experiments, we will focus on this aspect for the remaining part of this section. The current best upper bound on the $e$EDM, parametrized by $d_{e}$, given in Ref.~\cite{Roussy_2022}, and the projection of the ACME collaboration read~\cite{Kley_2021,Wu_2022} 
\begin{subequations}
\begin{align}
    \vert d_{e}/e \vert &< 4.1 \cdot 10^{-30} \cm \approx 2.1 \cdot 10^{-16} \GeV^{-1} \label{eq:EMDboundc}\\
    \nonumber\\
    \vert d_{e}^{\mathrm{ACME \, III}}/e \vert &< 0.3 \cdot 10^{-30} \cm \approx 1.5 \cdot 10^{-17} \GeV^{-1} \ .
\end{align}
\end{subequations}
Similarly, the current limit on the $\mu$EDM set by the muon~$\left( g-2 \right)$ experiment at Brookhaven National Laboratory and the projected ones by J-PARC and PSI~muEDM are~\cite{Bennett_2009,Abe_2019,Sakurai_2022}
\begin{subequations}
\begin{align}
    \vert d_{\mu}^{\mathrm{BNL}}/e \vert &< 1.9 \cdot 10^{-19} \cm \approx 9.6 \cdot 10^{-6} \GeV^{-1} \\
    \nonumber\\
    \vert d_{\mu}^{\mathrm{J-PARC}}/e \vert &< 1.5 \cdot 10^{-21} \cm \approx 7.6 \cdot 10^{-8} \GeV^{-1} \\
    \nonumber\\
    \vert d_{\mu}^{\mathrm{PSI}}/e \vert &< 6 \cdot 10^{-23} \cm \approx 3 \cdot 10^{-9} \GeV^{-1} \ .
\end{align}
\end{subequations}
As a consequence of the $\mathbb{Z}_{2}$ symmetry, the operator of the IDM EFT (\BSMEFT) contributes to the $\ell$EDM only at two-loop level at leading order, whereas the dominant contribution of the related SM EFT (\SMEFT) operator $\tilde{c}_{1} \left\vert H_{1} \right\vert^{2} V_{\mu \nu} \widetilde{V}^{\mu\nu}$ is a one-loop effect. The details of the calculation for both operators are presented in Appendix~\ref{App:CalculationeEDM}. The \SMEFT operator has been recently analyzed by Kley \emph{et al.}~\cite{Kley_2021} and it is considered here for the sake of comparison. Analogous to the analysis of the \BSMEFT operator, Fig.~\ref{Fig:EWBG_H1_asymmetryVSvev} illustrates the BAU obtained with the \SMEFT operator during the EWPhT associated with $H_{1}$.
As can be seen, a rather similar size of the $CP$-violating operator to the one involving the inert doublet, studied before, is required to arrive at the correct baryon abundance.

Choosing a rather generic value of $\tilde{c}_{1} = 15 \PeV^{-2}$ at the energy scale $\mu=m_{h}$, that reproduces the correct BAU, and utilizing the publicly available Mathematica package \DsixTools~\cite{Celis_2017,Fuentes_Mart_n_2021} for accounting of the running dictated by the renormalization group equations, the $\ell$EDM as derived in Sec.~\ref{Sec:SMEFToperator} reads 
\begin{align}
\label{eq:H1lEDM}
    \left\vert \frac{d_{\ell}^{H_{1}}}{e}\right\vert_{\mu=m_{\ell}} \approx \begin{cases} 2.6 \cdot 10^{-16} \GeV^{-1} \ \mathrm{for} \ \ell=e \\ 
    5.6 \cdot 10^{-14} \GeV^{-1} \ \mathrm{for} \ \ell=\mu\\ 
    9.7 \cdot 10^{-13} \GeV^{-1} \ \mathrm{for} \ \ell=\tau \end{cases} \ ,
\end{align} which is already in tension with the bound of Eq.~\eqref{eq:EMDboundc} (though it could still be met in corners of the parameter space).
In contrast, applying the result from Sec.~\ref{Sec:BSMoperator}, the $\ell$EDM $d_{\ell}^{H_{2}}$ induced by the \BSMEFT operator reads
\begin{align}
   \left\vert \frac{d_{\ell}^{H_{2}}}{e} \right\vert_{\mu=m_{\ell}} \approx \begin{cases} 5.9 \cdot 10^{-17}\GeV^{-1} \ \mathrm{for} \ \ell=e \\ 
    1.3 \cdot 10^{-14}\GeV^{-1} \ \mathrm{for} \ \ell=\mu \\ 
    2.2 \cdot 10^{-13}\GeV^{-1} \ \mathrm{for} \ \ell=\tau
    \end{cases} \, ,
\end{align}
where we have assumed a typical inert DM mass $m_{H}=71\GeV$ in the low-mass regime and the other inert states being degenerate in mass (throughout the paper), here with the splitting $\Delta m \equiv m_{H^{\pm},A} - m_H = 410\GeV$, and $\lambdaTFF=-0.002$, $\tilde{c}_{2}=25\PeV^{-2}$. These results suggest that the \BSMEFT operator can account for the BAU while generating an $e$EDM within the projected range of experimental sensitivity of ACME~III, however safely below the current limit. The EDMs of the other leptons are considerably out of reach.

Before closing the analysis of baryogenesis, it is worth pointing out potential improvements for a more accurate calculation of the BAU. 
For instance, in addition to investigating the impact of different bubble wall profiles on the effective chemical potential and thus on the maximally achievable BAU, a more precise description of the dynamics of the PhT, including the latent heat driving the expansion of the bubble and the frictional force the bubble experiences while expanding in the plasma, would allow to quantify the sphaleron dynamics and thereby the resultant BAU more accurately.

Finally, we would like to mention that the operator \eqref{eq:dual-dualOperator} is in fact quite unique when seeking to add $CP$ violation to the IDM involving the \emph{inert} Higgs. As demonstrated for this operator, but holding more generally, this has the advantage that EDMs arise at higher loops compared to the case of similar operators featuring $H_1$ -- the reason being that more lines, involving $H_2$ (that does not feature a zero-temperature vev), need to be closed. 

Potentially alternative choices for $CP$-violating terms involving $H_2$ read 
\begin{equation}
    {\cal L}_{\rm CPv}^{\rm alt} \supset\    
    C_{qH_{2}} H_1^\dagger H_2\,  \bar q_L H_2  q_R + C_{5} (H_1^\dagger H_2)^2 \left| H_1 \right|^2\,,
\end{equation} 
where in particular the first operator is interesting since it allows for a Yukawa-like interaction of the inert Higgs with fermions, respecting the $\mathbb{Z}_{2}$ symmetry. However, none of them is capable of injecting the sought $CP$ violation at the phase transition in the $H_2$ direction,
given that the background value of $H_1$, entering the operators, vanishes there -- and the same holds for the operators discussed below in Eq.~(\ref{eq:derivativeOperators}).

\section{Dark Matter results}
\label{Sec:DM}
In addition to the analysis on the possibility to attain the BAU in this model, it is also important to examine the impact on DM phenomenology. As found in the preceding section, the Wilson coefficient of the \BSMEFT operator must fulfil $\tilde{c}_{2} \sim 25\PeV^{-2}$ for a two-step EWPhT with a critical vev of $v_{c} \sim 100\GeV$ for generating a BAU matching the measured one. Here we discuss the consequences of this operator for DM physics. Therefore, we calculate the relic abundance as well as the direct-detection (DD) cross sections with the public \micromegas package~\cite{B_langer_2018}. The details of the analysis on the impact on the relic abundance and DD cross section are presented in Appendix~\ref{App:CrossSections_Impact}.

\begin{figure}[b!]
    \centering
    \includegraphics[width=0.75\textwidth]{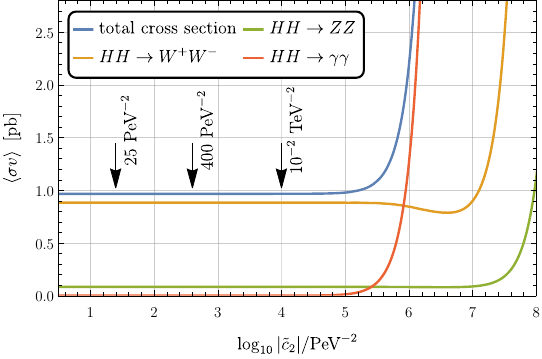}
    \caption{Thermally averaged annihilation cross section $\braket{\sigma v}$ for various channels in terms of the BSM coupling parameter $\tilde{c}_{2}$. The DM mass is $m_{H}=71\GeV$, the mass splitting $\Delta m = 410\GeV$, and the Higgs portal coupling was chosen to be $\lambdaTFF=-0.002$. The arrows indicate those values for $\tilde{c}_{2}$ considered in Fig.~\ref{Fig:EWBG_H2_asymmetryVSvev}.}
    \label{Fig:ThermalCrossSections_ctildeDependence_lowMass}
\end{figure}

\begin{figure}[b!]
    \centering
    \includegraphics[width=0.49\textwidth]{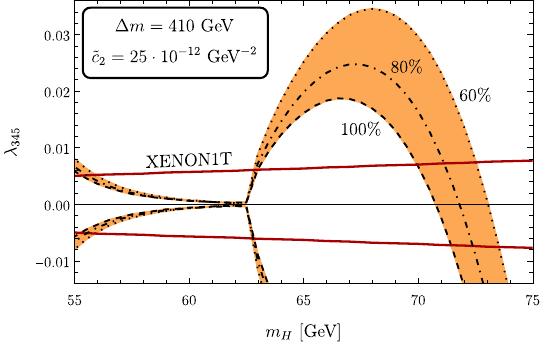}
    \includegraphics[width=0.49\textwidth]{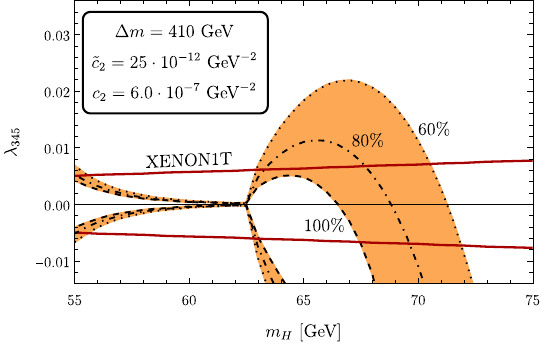}
    \caption{Parameter space for a sizable amount of the measured DM abundance in terms of the DM mass $m_{H}$ and the Higgs portal coupling $\lambdaTFF$ for fixed mass splitting $\Delta m$ and Wilson coefficients $\tilde{c}_{2}$, $c_{2}$ ($c_{2}=0$ if not specified otherwise). The inner boundary corresponds to the full relic abundance, \emph{i.e.} $\Omegahh = \Omegahh_{\mathrm{ref}}$, and the parameter space is truncated at $60\%$ of $\Omegahh_{\mathrm{ref}}$. The red solid lines enclose the region in agreement with XENON1T DD bounds. The dependence of the relic abundance on the mass splitting has been studied in Ref.~\cite{Fabian_2021}.}
    \label{Fig:LowMass_Relics_300_ctilde-12}
\end{figure}

Previous studies of the (original) IDM show that the interesting parameter space comprises the DM mass regimes of $55 \GeV \lesssim m_{H} \lesssim 80 \GeV$ and $m_{H}\gtrsim 500 \GeV$ (see, \emph{e.g.}, Refs.~\cite{Honorez_2007,Banerjee_2019}).\footnote{The first range can be even extended down to $\sim 44\GeV$ for a narrow BSM mass spectrum~\cite{Kalinowski_2021}.} In contrast to mass spectra with a large DM mass, the low-mass regime also features a suitable parameter space with a strong first-order EWPhT either via one step or two steps as described before. Therefore, together with the $CP$-violating operator, the low-mass regime can in principle accommodate DM and baryogenesis, provided that the impact of the new operator on the DM relic abundance is not harmful. First, we demonstrate in Fig.~\ref{Fig:ThermalCrossSections_ctildeDependence_lowMass} that the dimension-six operator contributes to the total thermally averaged annihilation cross section~$\braket{\sigma v}$ constructively, regardless of the sign of $\tilde{c}_{2}$.
As long as $\vert \tilde{c}_{2} \vert \lesssim 10^{-1} \TeV^{-2}$, the annihilation cross section $\braket{\sigma v}$ and thus the resultant relic abundance are virtually identical to the respective quantities in the vanilla IDM which means that the Wilson coefficients appearing in Fig.~\ref{Fig:EWBG_H2_asymmetryVSvev} clearly do not affect the DM relic abundance significantly. We emphasize that the $\tilde{c}_{2} = 25\PeV^{-2}$ does not change the relic abundance and yet delivers the measured BAU. Accordingly, the viable parameter space in the low-mass regime is shown in the left panel of Fig.~\ref{Fig:LowMass_Relics_300_ctilde-12}. 
Note that the mass splitting is sufficiently large, so that even larger mass splittings, as required for a two-step EWPhT, effectively do not change the surviving parameter set. The red lines represent the XENON1T DD bounds, indicating that only Higgs-portal couplings $\vert \lambda_{345} \vert \lesssim 0.01$ are experimentally allowed. 

Looking at Fig.~\ref{Fig:ThermalCrossSections_ctildeDependence_lowMass}, one can anticipate that increasing $\tilde{c}_{2}$ will lead to a shift of the viable colored area towards smaller $\lambda_{345}$ in the region of $m_H>m_h/2$. Interestingly, that could in principle enhance the possible DM parameter space, opening the region between $63$\,GeV and $70$\,GeV and thereby avoiding the necessity to sit in rather tuned regions, visible in the left plot of Fig.~\ref{Fig:LowMass_Relics_300_ctilde-12}. However, as it turns out, the corresponding required size of $\tilde{c}_{2}$ would lead to a significantly too large BAU. On the other hand, a UV completion that induces $\tilde{c}_{2} \vert H_{2} \vert^{2} V_{\mu \nu} \widetilde{V} ^{\mu \nu}$ is generically also expected to generate the $CP$-conserving operator $c_{2} |H_{2}|^2 V_{\mu\nu}V^{\mu\nu}$ (see Appendix~\ref{App:UVcompletion}), which does not impact the BAU. To explore this possibility, we show in Fig.~\ref{Fig:LowMass_Relics_300_ctilde-12} the corresponding DM parameter space for $c_{2}=6\cdot 10^{-7}\GeV^{-2}$.
We note that this would correspond to new particles not far above the TeV scale with ${\cal O}(1)$ $CP$-conserving couplings, while the respective $CP$-violating interactions would need to be some orders of magnitude smaller. Interestingly enough, there are completions where the $CP$-conserving operator receives additional contributions compared to the $CP$-violating one (see Appendix~\ref{App:UVcompletion} for more details). We now inspect that the formerly excluded parameter space opens and a viable DM abundance can be achieved for a much broader range of masses.
Fortunately, DD bounds are basically unaffected because the operator contributes to the DD cross section only at the loop level. In summary, our benchmark scenario provides successful baryogenesis together with a much broader range of viable DM masses of $55\,{\rm GeV} \lesssim m_H \lesssim 70\,{\rm GeV}$, compared to the original IDM.

\subsection*{Comments on High-Mass Regime}
The analysis of the extended IDM has shown so far that the $D=6$ operator can give rise to the measured DM relic abundance and the BAU with DM in the low-mass regime. In the remainder of this section we will pursue the question of whether corresponding parameter space exists also in the high-mass regime. Based on one of the findings in Ref.~\cite{Fabian_2021}, this regime does not feature a two-step EWPhT and hence renders the operator in Eq.~(\ref{eq:dual-dualOperator}) futile for producing the BAU. Yet, one can consider the $CP$-violating \SMEFT operator $\tilde{c}_{1} \vert H_{1} \vert^{2} V_{\mu \nu} \widetilde{V} ^{\mu \nu}$ for generating the BAU via a one-step EWPhT instead, see Fig.~\ref{Fig:EWBG_H1_asymmetryVSvev}. However, regardless of the choice of the two $D=6$ operators, the high-mass mass regime does not feature a strong first-order EWPhT while creating a substantial fraction of the DM relic abundance, as the latter requires a fairly degenerate BSM mass spectrum~\cite{Kalinowski_2021,Fabian_2021}. The reason for this is the increase of the  cross section of DM annihilation into longitudinal gauge bosons for larger mass splittings, \emph{i.e.} for $\Delta m \gtrsim 10 \GeV$ \cite{Fabian_2021}, which consequently results in underabundant DM. Nonetheless, it is precisely for $\Delta m \sim 200 \GeV$ that one can attain a strong first-order EWPhT in this regime. One way to potentially cure this problem is introducing further effective operators which modify  interactions between the DM particle and SM gauge bosons. The dimension-six operators that serve this purpose and that we will consider in the following read 
\begin{align}
     \mathcal{L}_{\mathrm{BSM}} &\supset C_{1} \left\vert H_{1} \right\vert^{2} \left( D_{\mu} H_{2} \right)^{\dagger} D^{\mu}H_{2} + C_{2} \left\vert H_{2} \right\vert^{2} \left( D_{\mu} H_{1} \right)^{\dagger} D^{\mu}H_{1}\nonumber\\
    &+ C_{3} \left[ H_{1}^{\dagger} H_{2} \left( D_{\mu} H_{1} \right)^{\dagger} D^{\mu}H_{2} + \mathrm{h.c.} \right] + C_{4} \left[ H_{1}^{\dagger}H_{2} \left( D_{\mu} H_{2} \right)^{\dagger} D^{\mu}H_{1} + \mathrm{h.c.} \right],
    \label{eq:derivativeOperators}
\end{align} 
where we take the four $C_{i}$ to be real\footnote{Note that $C_{3}$ and $C_{4}$ could be, in principle, complex. If that was the case, we would have additional sources of $CP$ violation that might have an impact on the BAU.} for the sake of simplicity. They are promising, since they can contribute to annihilation into longitudinal gauge bosons (\emph{i.e.}, Goldstone modes). The contributions of each of these operators to the total cross section are investigated in Appendix \ref{App:CrossSections_Impact}. We find that negative values of the Wilson coefficients lead to destructive interference and thus to an enhancement of the relic abundance. In fact, the behaviour of the total cross section is determined by an interplay between reducing the impact of the annihilations of two DM particles into EW gauge bosons and increasing the annihilations into a pair of either SM Higgs bosons or top quarks. A scan over possible values of the $C_i$ leads for example to a viable benchmark of
\begin{align}
   C_{1} = -5.4 \TeV^{-2} \dist C_{3} = -3.1 \TeV^{-2} \dist C_{4} = -3.2\TeV^{-2}
    \label{Eq:Wilsons}
    \end{align}
with $C_{2} \approx 0$. 

\begin{figure}[b!]
    \centering
    \includegraphics[width=0.49\textwidth]{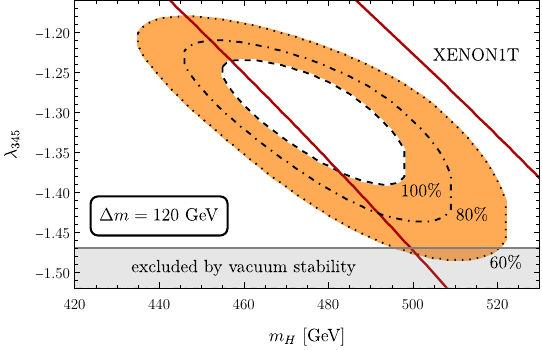}
    \includegraphics[width=0.49\textwidth]{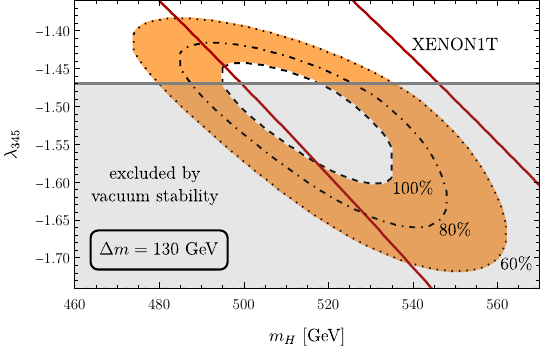}
    \caption{Relic abundance in terms of the DM mass $m_{H}$ and the Higgs portal coupling $\lambdaTFF$ for fixed Wilson coefficients (see Eq.~(\ref{Eq:Wilsons})) and two possible mass splittings $\Delta m$ with degenerate non-DM inert scalar masses. Since the $CP$-violating operator discussed in the previous section does not lead to a baryon asymmetry in the high-mass regime, it is turned off, \emph{i.e.} $\tilde{c}_{2}=0$. The parameter space is truncated at $60\%$ of $\Omegahh_{\mathrm{ref}}$ and the red lines represent the XENON1T DD bounds such that the parameter space between both lines is not excluded.}
    \label{Fig:HighMass-Relics}
\end{figure}

As can be seen in Fig.~\ref{Fig:HighMass-Relics}, this set allows to reproduce the measured DM relic abundance for a large mass splitting of $\Delta m \sim 120 \,{\rm GeV} \gg 10 \,{\rm GeV}$, while still respecting all experimental and theoretical constraints.
However, it turns out that this is not enough to reach a strong first-order EWPhT, in particular because also the large required $\vert \lambdaTFF \vert$ weakens the transition. Anyways, the extension of the viable DM region to significantly larger mass splitting furnishes already a significant first step towards a realistic model of baryogenesis and DM also in the high mass regime. In fact, further operators that are expected in typical UV completions (including those presented in the appendix), such as $\vert H_{1} \vert^{6}$, also enhance the EWPhT (see Refs.~\cite{Grojean:2004xa,Bodeker:2004ws,Delaunay:2007wb,Goertz:2015dba}) and a combined effect could lead to a strong transition. Still, regarding the beauty of minimality, the low mass regime arguably furnishes a more attractive scenario of baryogenesis and DM in the IDM framework.

\section{Conclusions}
\label{sec:Conc}
In this work, we investigated different effective operators to augment the IDM in order to fully account for baryogenesis without losing the DM candidate. We found that in the low-mass regime the \BSMEFT operator $\vert H_{2}\vert^{2} V_{\mu \nu} \widetilde{V}^{\mu \nu}$ allows to explain the measured BAU in addition to the DM abundance, indicating a beyond-IDM energy scale $\Lambda \sim 200\TeV$ (assuming an $\mathcal{O}(1)$ coupling) and avoiding stringent constraints from the $e$EDM (see Appendix~\ref{App:CalculationeEDM} for the details of the two-loop calculation). 
We also pointed out that once adding the corresponding $CP$-conserving operator, the viable DM range gets significantly broadened to $55\,{\rm GeV} \lesssim m_H \lesssim 70\,{\rm GeV}$.

On the contrary, the high-mass regime needs a few more effective operators due to the mutual exclusion of a sizable fraction of the DM relic abundance and an appropriate nature of the EWPhT in the original IDM. Considering the $CP$-violating \SMEFT operator $\vert H_{1}\vert^{2} V_{\mu \nu} \widetilde{V}^{\mu \nu}$ for generating the BAU indicates a scale $(\tilde c_1)^{-1/2} \sim 300\TeV$, while additional $D=6$ operators, detailed above, can help to reconcile the DM relic abundance and a strong EWPhT when appearing at a scale of $\mathcal{O}\left( 1\TeV \right)$.

In conclusion, our analysis demonstrates that the economic extension of the SM scalar sector by one inert $SU(2)_{L}$ doublet can in fact be a crucial first step towards a model that solves quantitatively some questions that the SM left open. Its augmentation with the advocated \BSMEFT operator delivers a simple and realistic benchmark that explains both the BAU and DM that can be investigated further. The EFT approach allows to cover a multitude of potential UV completions, with a couple of them being presented in Appendix~\ref{App:UVcompletion}.

\section*{Acknowledgements}
Many thanks go to Andrei Angelescu and Sudip Jana for fruitful discussions about the $\ell$EDM. Moreover, we are grateful to Elena Venturini, Jonathan Kley, Tobias Theil, and Andreas Weiler for helping us resolve a discrepancy regarding the calculation of the one-loop $\ell$EDM and to Matthias Neubert, Ulrich Nierste, and Tania Robens for useful remarks. Special thanks also to Jim Cline for helpful correspondence on details of the presented mechanism of baryogenesis. 

\clearpage

\begin{appendix}
\section{Calculation of the lepton EDM}\label{App:CalculationeEDM}
This appendix is dedicated to the explicit calculation of the EDM parameter~$d_{\ell}$ of the lepton~$\ell$ for the two $D=6$ operators involving the field strength tensors, \emph{i.e.} the one in Eq.~\eqref{eq:dual-dualOperator} and the similar operator featuring the SM-like Higgs instead of $H_{2}$. The low-energy effective operator associated with the $\ell$EDM reads
\begin{align}
    \mathcal{L}_{\mathrm{eff}}^{\mathrm{\ell EDM}} = -\frac{i}{2}d_{\ell}\ \bar{\ell} \sigma^{\mu\nu}\gamma_{5}\ell F_{\mu\nu} =  -\frac{i}{2}d_{\ell} \left( \overline{\ell_{L}} \sigma^{\mu\nu} \ell_{R} - \overline{\ell_{R}} \sigma^{\mu\nu} \ell_{L} \right) F_{\mu\nu}
    \label{Eq:eEDMLagrangian}
\end{align}
with $\sigma^{\mu\nu} \equiv i\left[\gamma^{\mu},\gamma^{\nu} \right]/2$ and the electromagnetic field strength tensor~$F_{\mu\nu}$ (see, \emph{e.g.}, Refs.~\cite{Aebischer_2021,Kley_2021}). The second expression contains the usual chiral projections $\ell_{L,R} \equiv (1\mp \gamma^{5})/2 \, \ell$.

The following calculation is based on `naive dimensional regularization', as discussed in Refs.~\cite{Buras_1998,Denner_2020,Kley_2021}, which retains the anti-commutation properties of $\gamma_5$ for any number of space-time dimensions.
The $\gamma_{5}$ matrix can be expressed in terms of the other $\gamma^{\mu}$ matrices and the Levi-Civita symbol $\varepsilon^{\mu\nu\rho\sigma}$ as $\gamma_{5} \equiv -i\varepsilon^{\mu\nu\rho\sigma}\gamma_{\mu}\gamma_{\nu}\gamma_{\rho}\gamma_{\sigma}/4!$ with $\varepsilon^{0123}=1$.
In the following, we consider a lepton with mass~$m_{\ell}$, electric charge~$Q_{\ell}$ in terms of the elementary charge~$e$, incoming momentum~$p_{1}$, and outgoing momentum~$k_{1}$, as well as an incoming photon with momentum~$p_{2}$.

\subsection{SM effective operator} \label{Sec:SMEFToperator}
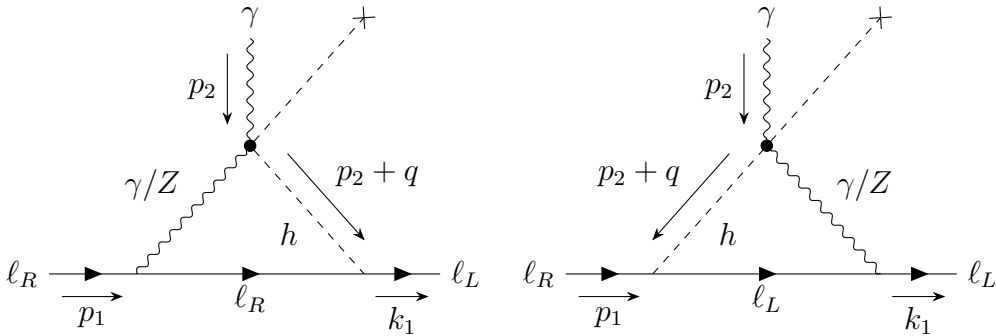
\begin{figure}[b]
\centering
\begin{tikzpicture}
\begin{feynman}
\vertex (a1) {\(\ell_{R}\)};
\vertex[right=1.5cm of a1] (a2);
\vertex[right=3cm of a2] (a3);
\vertex[right=1.0cm of a3] (a4) {\(\ell_{L}\)};

\vertex at ($(a2)!0.5!(a3) + (0, 1.7cm)$)[dot] (b1) {};

\vertex[above=1.7cm of b1] (c1) {\(\gamma\)};
\vertex at ($(a3) + (0,3.4cm)$) (c2);

\diagram* {
{[edges=fermion]
(a1) -- [momentum'=\(p_1\)] (a2) -- [edge label'=\(\ell_{R}\)] (a3) -- [momentum'=\(k_{1}\)] (a4),
},
(a2) -- [boson, edge label=\(\gamma/Z\)] (b1),
(b1) -- [scalar, edge label'=\(h\), momentum=\(p_{2}+q\)] (a3),

(b1) -- [boson, rmomentum=\(p_{2}\)] (c1),
(b1) -- [scalar, insertion=1] (c2)
};
\end{feynman}
\end{tikzpicture}
\hspace*{0mm}
\begin{tikzpicture}
\begin{feynman}
\vertex (a1) {\(\ell_{R}\)};
\vertex[right=1.5cm of a1] (a2);
\vertex[right=3cm of a2] (a3);
\vertex[right=1.0cm of a3] (a4) {\(\ell_{L}\)};

\vertex at ($(a2)!0.5!(a3) + (0, 1.7cm)$)[dot] (b1) {};

\vertex[above=1.7cm of b1] (c1) {\(\gamma\)};
\vertex at ($(a3) + (0,3.4cm)$) (c2);

\diagram* {
{[edges=fermion]
(a1) -- [momentum'=\(p_1\)] (a2) -- [edge label'=\(\ell_{L}\)] (a3) -- [momentum'=\(k_{1}\)] (a4),
},
(a2) -- [scalar, rmomentum=\(p_{2}+q\), edge label'=\(h\)] (b1),
(b1) -- [boson, edge label=\(\gamma/Z\)] (a3),

(b1) -- [boson, rmomentum=\(p_{2}\)] (c1),
(b1) -- [scalar, insertion=1] (c2)
};
\end{feynman}
\end{tikzpicture}
\caption{Feynman diagrams for processes contributing to $d_{\ell}$, including the momentum flow. The dotted vertices correspond to insertions of the $D=6$ operator and a cross attached to a dashed line indicates the SM Higgs vev entering the vertex factor. The right panel shows the respective `mirrored' diagram.}
\label{Fig:EDM_FeynmanDiagramsAppendix}
\end{figure}

Since the structure of the operator in Eq.~(\ref{Eq:eEDMLagrangian}) involves a chirality flip, the tree-level interaction between the photon and the lepton does not contribute to the $\ell$EDM in the model at hand. At leading order in perturbation theory the present $D=6$ operator connects the incoming photon via a loop (SM Higgs boson $h$ and photon $\gamma$ or $Z$ boson) with the lepton, as shown in Fig.~\ref{Fig:EDM_FeynmanDiagramsAppendix}.
Allowing for different coefficients for both field-strength terms in the following, \emph{i.e.} $\tilde{c}_{HW} \neq \tilde{c}_{HB}$, the $D=6$ operator becomes 
\begin{subequations}
\begin{align}
    &\hspace{35mm}\mathcal{L} \supset \vert H_{1} \vert^{2} \left( \tilde{c}_{HW} W_{\mu\nu}^{a} \widetilde{W}^{a, \mu\nu} + \tilde{c}_{HB} B_{\mu\nu}\widetilde{B}^{\mu\nu} \right) \label{Eq:SMEFToperators}\\
    \nonumber\\
    &\supset 2\varepsilon^{\mu\nu\rho\sigma} vh\partial_{\mu}A_{\nu} \left[ \left(\tilde{c}_{HW}\sin^{2}\theta_{W} + \tilde{c}_{HB}\cos^{2}\theta_{W}\right)\partial_{\rho}A_{\sigma} + \left(\tilde{c}_{HW}-\tilde{c}_{HB} \right)\sin2\theta_{W}\partial_{\rho}Z_{\sigma} \right] \label{Eq:SMEFToperatorsRecast}
\end{align}
\end{subequations}
and gives rise to those two Feynman diagrams, considered in this calculation for the $\ell$EDM for a general choice of the Wilson coefficients. For notational convenience, we define $\tilde{c}_{\gamma} \equiv \tilde{c}_{HW}\sin^{2}\theta_{W} + \tilde{c}_{HB}\cos^{2}\theta_{W}$ and $\tilde{c}_{Z} \equiv \tilde{c}_{HW}-\tilde{c}_{HB}$.

Focusing on the left-hand diagram in Fig.~\ref{Fig:EDM_FeynmanDiagramsAppendix} with a mediating photon and one specific chirality configuration, the matrix element reads 
\begin{align}
    i\mathcal{M}_{\gamma} = 4 m_{\ell} Q_{\ell}e \, \tilde{c}_{\gamma} \, \overline{u_{L}}(k_{1})  \int \frac{\dd^{d} q}{\left( 2\pi \right)^{d}} \frac{\left( \slashed{p_{1}} - \slashed{q} + m_{\ell} \right)\gamma_{\beta}q_{\alpha}\varepsilon^{\alpha\beta\kappa\nu}p_{2,\kappa} \epsilon_{\nu}(p_{2})}{\left[ \left( q + p_{2} \right)^{2} - m_{h}^{2} \right] q^{2} \left[ \left( q - p_{1} \right)^{2} - m_{\ell}^{2} \right]}  u_{R}(p_{1}) \,.
    \label{Eq:MatrixElementOneLoop}
\end{align}
Making use of the identity $\gamma_{\lambda}\gamma_{\beta} = \left( \{\gamma_{\lambda},\gamma_{\beta} \} + \left[ \gamma_{\lambda},\gamma_{\beta} \right] \right)/2 = g_{\lambda\beta} - i\sigma_{\lambda\beta}$, omitting the suppressed term proportional to the lepton mass in the numerator, and introducing the short-hand notation $\Xi_{a,b}\equiv (q+a)^{2} - b^{2}$ for the factors in the denominator coming from the propagators allow us to write 
\begin{align}
    i\mathcal{M}_{\gamma} \supset 4m_{\ell} Q_{\ell}e \, \tilde{c}_{\gamma} \, \overline{u_{L}}(k_{1})  \int_{q} \frac{\left( p_{1} - q \right)^{\lambda}q_{\alpha}}{\Xi_{p_{2},m_{h}} \Xi_{0,0} \Xi_{-p_{1},m_{\ell}}} \left( g_{\lambda\beta} - i\sigma_{\lambda\beta} \right) \varepsilon^{\alpha\beta\kappa\nu} u_{R}(p_{1}) p_{2,\kappa} \epsilon_{\nu}(p_{2}) \ . \label{Eq:EDM_SMEFT_integral}
\end{align}
As we will see later, the metric term does not contribute due to the anti-symmetry of the Levi-Civita tensor.
The integral in Eq.~(\ref{Eq:EDM_SMEFT_integral}) will appear frequently in the subsequent calculation and we will hence present its evaluation here. Recasting it by introducing the Feynman parameters $x,y,z$ leads to\footnote{For the sake of brevity, here and below the Dirac delta (here $\delta\left( x+y+z-1 \right)$) is included tacitly in the integral measure.} 
\begin{align}
    \int_{q} \frac{\left( p_{1} - q \right)^{\lambda}q_{\alpha}}{\Xi_{p_{2},m_{h}} \Xi_{0,0} \Xi_{-p_{1},m_{\ell}}}= 2 \int_{x,y,z}  \int_{\tilde{q}}\frac{\Theta_{\alpha}^{\lambda}}{\left( \tilde{q}^{2} - \Delta \right)^{3}}
\end{align}
with $\Theta_{\alpha}^{\lambda} \equiv \left( \left(1-y\right)p_{1} - \tilde{q} + xp_{2} \right)^{\lambda}\left( \tilde{q} - xp_{2} + yp_{1} \right)_{\alpha}$, and the shifted momentum $\tilde{q}$ and momentum-independent remainder $\Delta$ read (employing $p_1^2=k_1^2=m_\ell^2$ and $p_2^2=0$)
\begin{align}
    \tilde{q} \equiv q + xp_{2} - yp_{1} \dist \Delta \equiv xm_{h}^{2} + y^{2}m_{\ell}^{2} \ . \label{Eq:IDMEFT_momentumRedefinition}
\end{align}

As the denominator of the integrand is symmetric in the integration momentum~$\tilde{q}$ upon sign flip, terms of the numerator linear in $\tilde{q}$ will vanish after integration for symmetry reasons and only those terms containing either the product $\tilde{q}^{\lambda}\tilde{q}_{\alpha}$ or a $\tilde{q}$-independent numerator will remain. The former leads via dimensional regularization after a Wick rotation to Euclidean spacetime to
\begin{align}
    I^\lambda_\alpha \defEq \mu^{2\epsilon} \int \frac{\dd^{d} \tilde{q}}{\left( 2 \pi \right)^{d}} \frac{-\tilde{q}^{\lambda}\tilde{q}_{\alpha}}{\left( \tilde{q}^{2} - \Delta \right)^{3}} = -i \frac{\mu^{2\epsilon}}{d} \int \frac{\dd^{d} \tilde{q}_{E}}{\left( 2 \pi \right)^{d}} \frac{\tilde{q}_{E}^{2}g_{\alpha}^{\lambda}}{\left( \tilde{q}_{E}^{2} + \Delta \right)^{3}} = \frac{-i g_{\alpha}^{\lambda}\Gamma \left( \epsilon \right)}{4\left(4\pi\right)^{d/2}} \left( \frac{\mu^{2}}{\Delta} \right)^{\epsilon}
\end{align}
in $d=4-2\epsilon$ spacetime dimensions. The latter ($\tilde{q}$-independent numerator), on the other hand, becomes
\begin{align}
    I_{0} \defEq -i \mu^{2\epsilon} \int \frac{\dd^{d} \tilde{q}_{E}}{\left( 2 \pi \right)^{d}} \frac{C_{\alpha}^{\lambda} \left( x,y \right)}{\left( \tilde{q}_{E}^{2} + \Delta \right)^{3}} = -i \mu^{2\epsilon}\frac{C_{\alpha}^{\lambda} \left( x,y \right)}{\left(4\pi \right)^{d/2}}\frac{\Gamma\left( 1+\epsilon \right)}{\Gamma \left( 3 \right) \Delta^{1+\epsilon}}
\end{align}
with $C_{\alpha}^{\lambda} \left( x,y \right) \equiv \left( \left(1-y\right)p_{1} + xp_{2} \right)^{\lambda}\left(yp_{1} - xp_{2} \right)_{\alpha}$. Considering only the $\tilde{q}$-dependent numerator in the integrand, as the contributions from $I_{0}$ are further suppressed in $m_{\ell}^2/m_{h}^{2}\ll 1$, 
one gets in the $\overline{\mathrm{MS}}$ renormalization scheme
\begin{align}
    i \mathcal{M}_{\gamma} &\supset 8 m_{\ell} Q_{\ell}e \,\tilde{c}_{\gamma} \ \overline{u_{L}}(k_{1}) \int_{x,y,z} I^\lambda_\alpha \ \left( g_{\lambda\beta} - i\sigma_{\lambda\beta} \right)\varepsilon^{\alpha\beta\kappa\nu}u_{R}(p_{1}) p_{2,\kappa} \epsilon_{\nu}(p_{2}) \nonumber \\
    \nonumber \\
    &= - \frac{m_{\ell} Q_{\ell}e \tilde{c}_{\gamma}}{8\pi^{2}} \int_{x,y,z} \log \frac{\mu^{2}}{\Delta} \ \overline{u_{L}}(k_{1}) \sigma_{\alpha\beta}\varepsilon^{\alpha\beta\kappa\nu}u_{R}(p_{1}) p_{2,\kappa} \epsilon_{\nu}(p_{2}) +{\cal O}(\varepsilon)\ .
\end{align}
In practice, the divergence and constant term arising from the dimensional regularization are absorbed by the \SMEFT counterterm operator $\overline{\ell_{L}} \sigma^{\mu\nu} H_{1} e_{R} \left( B_{\mu\nu} + \sigma^{a}/2 \ W_{\mu\nu}^{a} \right)$ with the Pauli matrices~$\sigma^{a}$. Alternatively, one could just cut off the loop integral at the new-physics scale. 

Owing to the identities
\begin{align}
    \varepsilon^{\alpha\beta\kappa\nu}\sigma_{\alpha\beta} & \equiv -2i\gamma_{5}\sigma^{\kappa\nu}, \quad \left[\gamma_{5},\sigma^{\kappa\nu}\right]=0 \ , 
    \label{Eq:LeviCivitaGamma5Relation}
    \\[2mm]
    \qquad \sigma^{\mu\nu}F_{\mu\nu} & = \frac{i}{2} \left[ \gamma^{\mu}, \gamma^{\nu} \right] \left( \partial_{\mu}A_{\nu} - \partial_{\nu}A_{\mu} \right) = 2\sigma^{\mu\nu}\partial_{\mu}A_{\nu}
    \label{Eq:FieldStrengthTensorMomentumRelation} \ ,
\end{align}
the EDM parameter~$d_{\ell}^{(1)}$ for the first diagram can now be extracted from the matrix element as 
\begin{align}
    i \mathcal{M}_{\gamma} &\supset \underbrace{-\frac{m_{\ell} Q_{\ell}e \tilde{c}_{\gamma}}{4\pi^{2}} \  \int_{x,y,z} \log \frac{\mu^{2}}{\Delta}}_{= d_{\ell}^{(1)} \ \mathrm{from \ Eq.~(\ref{Eq:eEDMLagrangian})}} \ \overline{u_{L}}(k_{1}) \sigma^{\kappa\nu} \underbrace{\gamma_{5}u_{R}(p_{1})}_{=u_{R}(p_{1})} \left( -i p_{2,\kappa} \right) \epsilon_{\nu}(p_{2})\,,
\end{align}
and reads
\begin{align}
    \left(\frac{d_{\ell}^{(1)}}{e}\right)_{\gamma} = -\frac{m_{\ell} Q_{\ell} \tilde{c}_{\gamma}}{4\pi^{2}} \ \int_{0}^{1}\dd x \int_{0}^{1-x} \dd y \ \log \frac{\mu^{2}}{xm_{h}^{2} + y^{2}m_{\ell}^{2}}\,.
\end{align}
The contribution $d_{\ell}^{(2)}$ of the `mirrored' diagram (see Fig.~\ref{Fig:EDM_FeynmanDiagramsAppendix}) equals the first one. Hence, expanding in $m_{\ell} \ll m_{h}$ ultimately leads to the $\ell$EDM, reading
\begin{align}
    \left(\frac{d_{\ell}}{e}\right)_{\gamma} = \left(\frac{d_{\ell}^{(1)}}{e}\right)_{\gamma} + \left(\frac{d_{\ell}^{(2)}}{e}\right)_{\gamma} = -m_{\ell} Q_{\ell} \left( \frac{3}{8\pi^{2}}  + \frac{1}{2\pi^{2}}\log\frac{\mu}{m_{h}} \right)\tilde{c}_{\gamma} \, .
\end{align}

Considering a mediating $Z$ boson, coupling to the right-handed lepton in the first diagram, the matrix element reads
\begin{align}
    i&\mathcal{M}_{Z} = 2m_{\ell} \frac{gc_{R}}{\cos\theta_{W}}\tilde{c}_{Z}\sin2\theta_{W} \int_{q} \frac{\overline{u_{L}}\left( k_{1} \right) \left( \slashed{p_{1}} - \slashed{q} + m_{\ell} \right) \gamma_{\beta} q_{\alpha} \varepsilon^{\alpha\beta\kappa\nu} u_{R}\left( p_{1} \right) p_{2,\kappa}\epsilon_{\nu}\left( p_{2} \right)}{\left[ \left( q + p_{2} \right)^{2} - m_{h}^{2} \right] \left(q^{2} - m_{Z}^{2} \right) \left[ \left( q - p_{1} \right)^{2} - m_{\ell}^{2} \right]}\nonumber\\
    \nonumber\\
    &\supset \underbrace{-\frac{m_{\ell} c_{R}e\tilde{c}_{Z}}{4\pi^{2}} \int_{x,y,z} \log\frac{\mu^{2}}{xm_{h}^{2} + zm_{Z}^{2}}}_{= d_{\ell}^{(1)} \ \mathrm{from \ Eq.~(\ref{Eq:eEDMLagrangian})}} \ \overline{u_{L}} \left( k_{1} \right) \gamma_{5}\sigma^{\kappa\nu} u_{R}\left( p_{1} \right) \left( -i p_{2,\kappa} \right) \epsilon_{\nu} \left( p_{2} \right) \label{Eq:IntegralZboson} \, ,
\end{align}
where we have, as before, introduced Feynman parameters, performed a Wick rotation, neglected the lepton mass, and considered the term proportional to the squared momentum in the integral (as shown explicitly above). Evaluating the integral for the massive mediator, 
\begin{align}
    \int_{0}^{1}\dd x\int_{0}^{1-x}\dd z \ \log\frac{\mu^{2}}{xm_{h}^{2} + zm_{Z}^{2}} &= \frac{3}{4} + \frac{m_{Z}^{4}\log\frac{\mu^{2}}{m_{Z}^{2}} - 2m_{h}^{2}m_{Z}^{2}\log\frac{\mu^{2}}{m_{Z}m_{h}} + m_{h}^{4}\log\frac{\mu^{2}}{m_{h}^{2}}}{2\left( m_{h}^{2} - m_{Z}^{2} \right)^{2}}\nonumber\\
    \nonumber\\
    &= \frac{3}{4} + \frac{\left( m_{Z}^{2} - m_{h}^{2} \right)^{2} \log\frac{\mu^{2}}{m_{h}^{2}} + 2m_{Z}^{2}\left( m_{Z}^{2} - m_{h}^{2} \right) \log\frac{m_{h}}{m_{Z}}}{2\left( m_{h}^{2} - m_{Z}^{2} \right)^{2}}\nonumber\\
    \nonumber\\
    &= \frac{3}{4} + \log\frac{\mu}{m_{h}} + \frac{m_{Z}^{2}}{m_{Z}^{2} - m_{h}^{2}}\log\frac{m_{h}}{m_{Z}} \ ,
\end{align}
and taking the `mirrored' diagram into account, gives rise to
\begin{align}
    \left(\frac{d_{\ell}}{e}\right)_{Z} &= -m_{\ell} \left[ \frac{c_{L} + c_{R}}{4\pi^{2}} \int_{x,y,z} \log\frac{\mu^{2}}{xm_{h}^{2} + zm_{Z}^{2}}\right]\tilde{c}_{Z}\nonumber \\
    \nonumber\\
    &=  -m_{\ell} \left[ \frac{3 \left(T_{3,\ell} - 2Q_{\ell}\sin^{2}\theta_{W}\right)}{16\pi^{2}} + \frac{T_{3,\ell} - 2Q_{\ell}\sin^{2}\theta_{W}}{4\pi^{2}} \log\frac{\mu}{m_{h}}\right.\nonumber\\
    &\hspace*{25mm}\left.+ \frac{T_{3,\ell} - 2Q_{\ell}\sin^{2}\theta_{W}}{4\pi^{2}} \frac{m_{Z}^{2}}{m_{Z}^{2} - m_{h}^{2}} \log\frac{m_{h}}{m_{Z}} \right]\tilde{c}_{Z}\,,
\end{align}
where we have used $c_{L} + c_{R} = T_{3,\ell} - 2Q_{\ell}\sin^{2}\theta_{W}$. Consequently, the full $\ell$EDM ultimately amounts to 
\begin{subequations}
\begin{align}
    &\frac{d_{\ell}}{e} = -m_{\ell} \left[\frac{3 T_{3,\ell}}{16\pi^{2}} + \frac{T_{3,\ell}}{4\pi^{2}} \log\frac{\mu}{m_{h}} + \frac{\left(T_{3,\ell} - 2Q_{\ell}\sin^{2}\theta_{W}\right)m_{Z}^{2}}{4\pi^{2}\left( m_{Z}^{2} - m_{h}^{2} \right)}  \log\frac{m_{h}}{m_{Z}} \right] \tilde{c}_{HW}\\
    \nonumber\\
    &-m_{\ell} \left[\frac{3 \left(2Q_{\ell} - T_{3,\ell}\right)}{16\pi^{2}} + \frac{2Q_{\ell}- T_{3,\ell}}{4\pi^{2}} \log\frac{\mu}{m_{h}} - \frac{\left(T_{3,\ell} - 2Q_{\ell}\sin^{2}\theta_{W}\right) m_{Z}^{2}}{4\pi^{2}\left( m_{Z}^{2} - m_{h}^{2} \right)} \log\frac{m_{h}}{m_{Z}} \right]\tilde{c}_{HB}
\end{align}
\end{subequations}
and matches\footnote{A missing factor of 2 has been corrected in a revised version of Ref.~\cite{Kley_2021}.} the one by Kley \emph{et al.}~\cite{Kley_2021} with $\tilde{c}_{HB} = \tilde{c}_{HW} = \tilde{c}_{1}$, reading
\begin{align}
    \frac{d_{\ell}}{e} = -\frac{m_{\ell}Q_{\ell}}{8\pi^{2}} \left( 3 + 4 \log \frac{\mu}{m_{h}} \right)\tilde{c}_{1} \ . \label{Eq:lEDMresultH1}
\end{align}

\subsection{IDM effective operator} 
\label{Sec:BSMoperator}
In addition to the \SMEFT operator in Eq.~(\ref{Eq:SMEFToperators}), the IDM effective field theory (\BSMEFT) operator reads
\begin{subequations}
\begin{align}
    &\hspace{30 mm}\mathcal{L} \supset \vert H_{2} \vert^{2} \left( \tilde{c}_{HW}^{\prime}W_{\mu\nu}^{a} \widetilde{W}^{a, \mu\nu} + \tilde{c}_{HB}^{\prime} B_{\mu\nu}\widetilde{B}^{\mu\nu} \right) \label{Eq:BSMEFToperators} \\
    \nonumber\\
    & \supset H^{2} \partial_{\mu}A_{\nu}\left[\left( \tilde{c}_{HW}^{\prime} \sin^{2}\theta_{W} + \tilde{c}_{HB}^{\prime} \cos^{2}\theta_{W}\right)\partial_{\rho}A_{\sigma} + \left( \tilde{c}_{HW}^{\prime} - \tilde{c}_{HB}^{\prime} \right) \sin 2\theta_{W} \partial_{\rho}Z_{\sigma}\right]\varepsilon^{\mu\nu\rho\sigma} \, . \label{Eq:BSMEFToperatorsRecast}
\end{align}
\end{subequations}
As the BSM Higgs doublet does not acquire a vev at zero temperature, the contribution to the $\ell$EDM occurs at two-loop level for the first time: a loop involving $H$, $A$, or $H^{\pm}$ connects the effective vertex to the SM Higgs boson. With the assignment of the particles' momenta given in the left-hand Feynman diagram in Fig.~\ref{Fig:EDM_FeynmanDiagrams2LoopAppendix}, the corresponding matrix element for a mediating photon and an $H$-loop for instance reads 
\begin{align}
    &i\mathcal{M}_{\gamma} = \frac{4}{S} i m_{\ell}Q_{\ell}e  \tilde{c}_{\gamma}^{\prime}\lambdaTFF \ p_{2,\rho} \epsilon_{\sigma} \left( p_{2} \right) \nonumber\\
    &\hspace{-3mm} \times \underbrace{\int_{q_{1}} \frac{\overline{u_{L}}\left( k_{1} \right) \left( \slashed{p_{1}} - \slashed{q_{1}} + m_{\ell} \right)\gamma_{\nu}q_{1,\mu}\varepsilon^{\mu\nu\rho\sigma} u_{R}\left(p_{1}\right)}{\left[ \left( p_{1} - q_{1}\right)^{2} - m_{\ell}^{2} \right] \left[ \left( q_{1} + p_{2} \right)^{2} - m_{h}^{2} \right] q_{1}^{2}}}_{\defEq \overline{u_{L}}\left( k_{1} \right) I_{1}^{\rho\sigma} u_{R}\left(p_{1}\right)} \underbrace{\int_{q_{2}} \frac{1}{\left[ \left( q_{1} - q_{2} + p_{2} \right)^{2} - m_{H}^{2} \right] \left[ q_{2}^{2} - m_{H}^{2} \right]}}_{\defEq I_{2}} \label{Eq:MatrixElementBSMEFT}
\end{align}
\begin{figure}[b!]
\centering
\begin{tikzpicture}
\begin{feynman}
\vertex (a1) {\(\ell_{R}\)};
\vertex[right=1.5cm of a1] (a2);
\vertex[right=3cm of a2] (a3);
\vertex[right=1.0cm of a3] (a4) {\(\ell_{L}\)};

\vertex at ($(a2)!0.25!(a3) + (0, 1.7cm)$)[dot] (b1) {};
\vertex at ($(a2)!0.75!(a3) + (0, 1.7cm)$) (b2);
    
\vertex[above=1.7cm of b1] (c1) {\(\gamma\)};
\vertex at ($(a3) + (0, 3.4cm)$) (c2);

\diagram* {
{[edges=fermion]
(a1) -- [momentum'=\(p_{1}\)] (a2) -- [edge label'=\(\ell_{R}\)] (a3) -- [momentum'=\(k_{1}\)] (a4),
},
(a2) -- [boson, momentum=\(q_{1}\)] (b1),
(b1) -- [scalar, edge label=\(H/A\), out=90, in=90, looseness=1.5] (b2),
(b1) -- [scalar, out=-90, in=-90, looseness=1.5, momentum'={[arrow shorten=0.3]\(q_{2}\)}] (b2),
(b2) -- [scalar, edge label=\(h\)] (a3),

(b1) -- [boson, rmomentum=\(p_{2}\)] (c1),
(b2) -- [scalar, insertion=1] (c2)
};
\end{feynman}
\end{tikzpicture}
\hspace*{0mm}
\begin{tikzpicture}
\begin{feynman}
\vertex (a1) {\(\ell_{R}\)};
\vertex[right=1.5cm of a1] (a2);
\vertex[right=3cm of a2] (a3);
\vertex[right=1.0cm of a3] (a4) {\(\ell_{L}\)};

\vertex at ($(a2)!0.25!(a3) + (0, 1.7cm)$)[dot] (b1) {};
\vertex at ($(a2)!0.75!(a3) + (0, 1.7cm)$) (b2);
    
\vertex[above=1.7cm of b1] (c1) {\(\gamma\)};
\vertex at ($(a3) + (0, 3.4cm)$) (c2);

\diagram* {
{[edges=fermion]
(a1) -- [momentum'=\(p_{1}\)] (a2) -- [edge label'=\(\ell_{L}\)] (a3) -- [momentum'=\(k_{1}\)] (a4),
},
(a2) -- [boson, momentum=\(q_{1}\)] (b1),
(b1) -- [anti charged scalar, edge label=\(H^{\pm}\), out=90, in=90, looseness=1.5] (b2),
(b1) -- [charged scalar, out=-90, in=-90, looseness=1.5, momentum'={[arrow shorten=0.3]\(q_{2}\)}] (b2),
(b2) -- [scalar, edge label=\(h\)] (a3),

(b1) -- [boson, rmomentum=\(p_{2}\)] (c1),
(b2) -- [scalar, insertion=1] (c2)
};
\end{feynman}
\end{tikzpicture}
\caption{Two-loop contribution to the $\ell$EDM from the BSM operator $\vert H_{2} \vert^{2} F_{\mu\nu}\widetilde{F}^{\mu\nu}$. The incoming vector boson is (by definition) a photon, but the internal one can be either a photon or a $Z$ boson.}
\label{Fig:EDM_FeynmanDiagrams2LoopAppendix}
\end{figure}
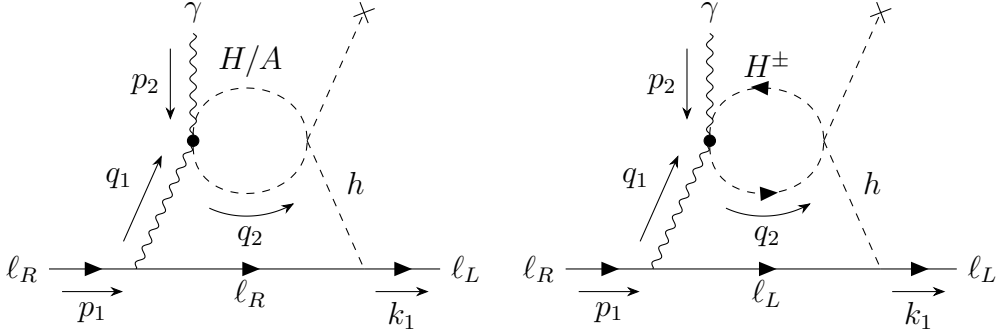with the symmetry factor $S$ (here $S=2$) and the definition $\tilde{c}_{\gamma}^{\prime} \equiv \tilde{c}_{HW}^{\prime}\sin^{2}\theta_{W} + \tilde{c}_{HB}^{\prime}\cos^{2}\theta_{W}$. 

After introducing the Feynman parameter $x_{2}$, the second integral becomes
\begin{align}
    I_{2} = \int_{\tilde{q}_{2}}\int_{x_{2}} \frac{1}{\left( \tilde{q}_{2}^{2} - \Delta_{2} \right)^{2}}
\end{align}
with the shifted momentum $\tilde{q}_{2} = q_{2} - x_{2}\left( q_{1} + p_{2} \right)$ and $\Delta_{2} = m_{H}^{2} - x_{2}\left( 1 - x_{2} \right) \left( q_{1} + p_{2} \right)^{2}$. Performing the Wick rotation leads to
\begin{align}
    I_{2} &= i \int_{x_{2}}\int_{\tilde{q}_{2,E}} \frac{1}{\left( \tilde{q}_{2,E}^{2} + \Delta_{2} \right)^{2}} = i \frac{\Gamma \left( \epsilon \right)}{\left( 4\pi \right)^{d/2}} \int_{x_{2}} \left( \frac{\mu^{2}}{\Delta_{2}} \right)^{\epsilon} \ . \label{Eq:IntegralIndex2}
\end{align}
Introducing the Feynman parameters $y_{1}, z_{1}$ for the first integral, the product of integrals becomes
\begin{align}
    I_{1}^{\rho\sigma}I_{2} &= 2\int_{y_{1},z_{1}}\int_{\tilde{q}_{1}} \ \frac{\left[ \left( 1-y_{1} \right)\slashed{p_{1}} - \slashed{\tilde{q}_{1}} + z_{1}\slashed{p_{2}} + m_{\ell} \right] \gamma_{\nu} \left( \tilde{q}_{1} + y_{1}p_{1} - z_{1}p_{2} \right)_{\mu} \varepsilon^{\mu\nu\rho\sigma}}{\left( \tilde{q}_{1}^{2} - \Delta_{1} \right)^{3}} I_{2}\left( \tilde{q}_{1} \right)
\end{align}
with $\tilde{q}_{1} = q_{1} - y_{1}p_{1} + z_{1}p_{2}$ and $\Delta_{1} = z_{1}m_{h}^{2} + y_{1}^{2}m_{\ell}^{2}$. 
Analogously to the calculation in Sec.~\ref{Sec:SMEFToperator}, we keep only the leading term in the numerator that is quadratic in $\tilde{q}_{1}$ and thus find 
\begin{align}
    I_{1}^{\rho\sigma}I_{2} \supset \frac{-2i\Gamma \left( \epsilon \right)}{\left( 4\pi \right)^{d/2}} \int\hspace{-3mm}\int_{\tilde{q}_{1}} \left( \frac{\mu^{2}}{m_{H}^{2} - x_{2} \left( 1-x_{2} \right)\left( \tilde{q}_{1} + y_{1}p_{1} + \left( 1-z_{1} \right)p_{2} \right)^{2}} \right)^{\epsilon} \frac{\tilde{q}_{1}^{\lambda}\tilde{q}_{1,\mu}\varepsilon^{\mu\nu\rho\sigma}\gamma_{\lambda}\gamma_{\nu}}{\left( \tilde{q}_{1}^{2} - \Delta_{1} \right)^{3}}
\end{align}
with the first integral being over the Feynman parameters $x_{2},y_{1},z_{1}$. Assuming negligibly small ratios $p_{1,2}/m_{H}$ gives rise to 
\begin{align}
    I_{1}^{\rho\sigma}I_{2} &\supset  \frac{-2i\Gamma \left( \epsilon \right)}{\left( 4\pi \right)^{d/2}} \left( \frac{\mu^{2}}{m_{H}^{2}} \right)^{\epsilon} \int\hspace{-3mm}\int_{\tilde{q}_{1}} \left( 1-x_{2}\left( 1 - x_{2} \right) \frac{\tilde{q}_{1}^{2}}{m_{H}^{2}} \right)^{-\epsilon} \frac{\tilde{q}_{1}^{\lambda}\tilde{q}_{1,\mu}\varepsilon^{\mu\nu\rho\sigma}\gamma_{\lambda}\gamma_{\nu}}{\left( \tilde{q}_{1}^{2} - \Delta_{1} \right)^{3}}\\
    \nonumber \\
    &= \frac{4i\Gamma \left( \epsilon \right)}{\left( 4\pi \right)^{d/2}d} \left( \frac{\mu^{2}}{m_{H}^{2}} \right)^{\epsilon} \int\hspace{-3mm}\int_{\tilde{q}_{1}} \left( 1-x_{2}\left( 1 - x_{2} \right) \frac{\tilde{q}_{1}^{2}}{m_{H}^{2}} \right)^{-\epsilon} \frac{\tilde{q}_{1}^{2}}{\left( \tilde{q}_{1}^{2} - \Delta_{1} \right)^{3}} \gamma_{5}\sigma^{\rho\sigma} \ ,
\end{align}
 where we applied the relation in Eq.~(\ref{Eq:LeviCivitaGamma5Relation}). 
This integral can be evaluated in Euclidean space and reads
\begin{align}
    I_{1}^{\rho\sigma}I_{2} &\supset \int_{x_{2},y_{1},z_{1}} \left( \frac{-\frac{2}{\epsilon} +3+4\gamma_{E} -2\log\frac{16\pi^{2}\mu^{4}\Delta_{1}^{2}}{x_{2}\left( 1 - x_{2} \right)m_{H}^{8}}}{1024\pi^{4} \epsilon} + F(m_H) + \mathcal{O}\left( \epsilon \right) \right) \gamma_{5}\sigma^{\rho\sigma} \ .
\end{align}
The present divergences can be eliminated by introducing appropriate counterterms $\overline{\ell_{L}} \sigma^{\mu\nu} H_{1} e_{R} \left( B_{\mu\nu} + \sigma^{a}/2 \ W_{\mu\nu}^{a} \right) + \mathrm{h.c.}$ as in the previous section, so that we can focus solely on the mass-dependent finite part $F(m )$ of the integral, which is rather lengthy and thus not displayed here. 
The corresponding matrix element reads
\begin{align}
    i\mathcal{M} \supset \underbrace{-\frac{4}{S} m_{\ell}Q_{\ell}e  \tilde{c}_{\gamma}^{\prime}\lambda_{345}\int_{x_{2},y_{1},z_{1}} F\left( m \right)}_{=d_{\ell}} \ \overline{u_{L}} \left( k_{1} \right) \gamma_{5}\sigma^{\mu\nu} u_{R}\left( p_{1} \right)\left( -i p_{2,\mu} \right) \epsilon_{\nu} \left( p_{2} \right) \label{Eq:EDMmatching}
\end{align}
with the SMEFT Wilson coefficient $\tilde{c}_{\gamma}^{\prime}$ defined at the scale $m_H$, being agnostic about its nature at this point. Note that we assume corrections involving lepton masses to be negligible, as they are considerably lighter than the (B)SM Higgs.

Taking the contributions of $H$, $A$, and $H^{\pm}$ into account, together with their respective symmetry factors ($S=2$ for $H$, $A$; $S=1$ for $H^{\pm}$), we find with degenerate BSM non-DM fields for the $\ell$EDM parameter
\begin{align}
   \frac{d_{\ell}}{e} &= -2\left[ \lambdaTFF \int F\left( m_{H} \right) + \left( \bar{\lambda}_{345} + \lambdaThree\right) \int F\left( m_{A,H^{\pm}} \right) \right] m_{\ell}Q_{\ell} \tilde{c}_{\gamma}^{\prime} \nonumber\\
   \nonumber\\
   &= -2m_{\ell}Q_{\ell} \tilde{c}_{\gamma}^{\prime} \left( \lambdaTFF \int \left[ F\left( m_{H} \right) + 2 F\left( m_{A,H^{\pm}} \right) \right] + 4 \frac{m_{A,H^{\pm}}^{2} - m_{H}^{2}}{v^{2}} \int F\left( m_{A,H^{\pm}} \right) \right) \, .
\end{align}
Since we chose $\tilde{c}_{HW}^{\prime}=\tilde{c}_{HB}^{\prime}$ for simplicity, where the $Z$ contribution to the $\ell$EDM vanishes, we do not derive the $Z$ contribution in the \BSMEFT. Considering the parameters $\mu=m_{H}=71\GeV$, the degenerate non-DM masses $m_{A,H^{\pm}}=481\GeV$, $\lambdaTFF=-0.002$, and the Wilson coefficient~$\tilde{c}_{2} \equiv \tilde{c}_{\gamma}^{\prime} = 25\PeV^{-2}$, we find after numerical integration
\begin{align}
    \left\vert \frac{d_{\ell}}{e} \right\vert_{\mu=m_{H}} \approx \begin{cases} 6.7 \cdot 10^{-17}\GeV^{-1} \ \mathrm{for} \ \ell=e \\ 
    1.4 \cdot 10^{-14}\GeV^{-1} \ \mathrm{for} \ \ell=\mu \\ 
    2.3 \cdot 10^{-13}\GeV^{-1} \ \mathrm{for} \ \ell=\tau
    \end{cases} \, ,
\end{align}
and running the $\ell$EDM parameter down to $\mu=m_{\ell}$ results in
\begin{align}
    \left\vert \frac{d_{\ell}}{e} \right\vert_{\mu=m_{\ell}} \approx \begin{cases} 5.9 \cdot 10^{-17}\GeV^{-1} \ \mathrm{for} \ \ell=e \\ 
    1.3 \cdot 10^{-14}\GeV^{-1} \ \mathrm{for} \ \ell=\mu \\ 
    2.2 \cdot 10^{-13}\GeV^{-1} \ \mathrm{for} \ \ell=\tau
    \end{cases} \, .
\end{align}

\clearpage

\section{Contributions of the Covariant-Derivative operators to the Dark-Matter cross sections}
\label{App:CrossSections_Impact}

In the following, we will consider the dimension-six operators of \eqref{eq:derivativeOperators}, containing the SM gauge covariant derivative. As the center-of-mass energy of the annihilating DM particles is much larger than the masses of the SM gauge bosons involved, we apply the Goldstone boson equivalence theorem and therefore consider only the longitudinally polarized gauge bosons in the respective final states.

Since \calchep does not take more than four fields for the computation of the cross sections into account, we keep only those -- most important -- terms in the following discussion.

\subsection{Impact on thermally averaged annihilation cross section $\braket{\sigma v}$}
\begin{figure}[b!]
    \centering
    \includegraphics[width=0.48\textwidth]{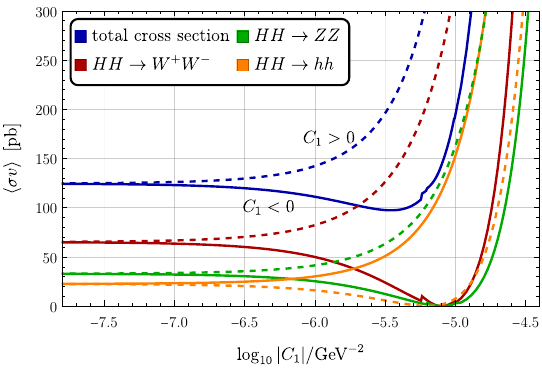}
    \includegraphics[width=0.48\textwidth]{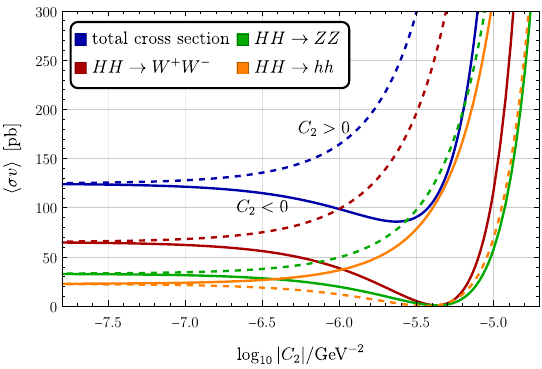}
    \caption{Evolution of the thermally averaged cross section $\braket{\sigma v}$ with respect to $C_{1}$ (left) and $C_{2}$ (right). The DM mass is $m_{H}=490\GeV$, the mass splitting between the degenerate inert scalars and the DM particle is $\Delta m = 120\GeV$, and the Higgs portal coupling reads $\lambdaTFF=-1.3$.}
    \label{Fig:ThermalCrossSection_c12Dependence}
\end{figure}

\begin{figure}[b!]
\centering
\includegraphics[width=0.48\textwidth]{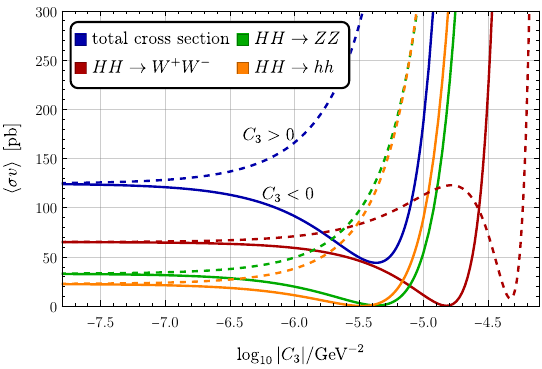}
\includegraphics[width=0.48\textwidth]{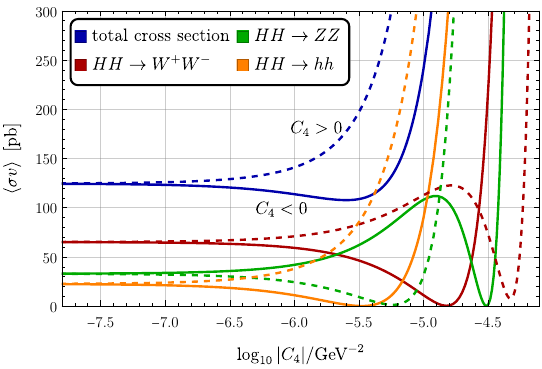}
\caption{Evolution of the thermally averaged cross section with respect to $C_{3}$ (left) and $C_{4}$ (right). The model parameters are the same as for Fig.~\ref{Fig:ThermalCrossSection_c12Dependence}.}
\label{Fig:ThermalCrossSection_c34Dependence}
\end{figure}

The first operator, associated with the Wilson coefficient $C_{1}$, induces
\begin{align}
    \left\vert H_{1} \right\vert^{2} \left\vert D_{\mu} H_2 \right\vert ^2 &\supset  \frac{1}{4}\partial_{\mu}H\partial^{\mu}H \left( h^{2} + 2vh + G^{2} + 2 G^{+}G^{-}\right)\,,
    \label{Eq:Vertex_c1}
\end{align}
and the evolution of the thermally averaged annihilation cross section with respect to $C_1$, as well as the contributions of the most relevant and interesting annihilation processes, are visualized in the left panel of Fig.~\ref{Fig:ThermalCrossSection_c12Dependence}. As expected for interference effects in the calculation of the cross sections, the sign of the Wilson coefficient significantly affects the annihilation cross sections for large values. In turn, the difference between the cross sections for opposite signs tends to zero as the effect becomes marginal for sufficiently small Wilson coefficients and the curve approaches the annihilation cross section governed by the renormalizable (vanilla) IDM. This expected effect is evident in each plot of Figs.~\ref{Fig:ThermalCrossSection_c12Dependence}-\ref{Fig:ThermalCrossSection_c34Dependence}.

The second dimension-six operator that we consider leads to
\begin{align}
\left\vert H_{2} \right\vert^{2} \left\vert D_{\mu} H_1 \right\vert^{2} &\supset \frac{1}{4}H^{2}\left( \partial_{\mu}h\partial^{\mu}h + \partial_{\mu}G\partial^{\mu}G + 2\partial_{\mu}G^{-}\partial^{\mu}G^{+} \right) \, .
\label{Eq:Vertex_c2}
\end{align}
Its impact on the DM annihilation cross section is depicted in the right panel of Fig.~\ref{Fig:ThermalCrossSection_c12Dependence}. The behaviour of the four processes is qualitatively the same for both effective operators. The minima of the annihilation cross sections for $HH\rightarrow ZZ$ and $HH\rightarrow W^{+}W^{-}$ are located at the same value of the Wilson coefficient, since the contributions of the annihilation channels (\emph{i.e.} four-point interaction, $s$- and $t$-channel, and $u$-channel if necessary) for both processes are equal pairwise. 

The third and fourth operator, corresponding to $C_{3}$ and $C_{4}$, respectively, lead to
\begin{align}
    H_{1}^{\dagger} H_{2} \left( D_{\mu} H_{1} \right)^{\dagger} D^{\mu}H_{2} + \mathrm{h.c.} &\supset \frac{1}{2}\partial_{\mu}H \left( H h \partial^{\mu}h + v H \partial^{\mu}h  - H G \partial^{\mu}G + v A \partial^{\mu}G \right) \nonumber\\
    &\hspace{5mm} + \frac{v}{2} H \left( \partial_{\mu}G \partial^{\mu}A + \partial_{\mu}G^{-} \partial^{\mu}H^{+} + \partial_{\mu}G^{+} \partial^{\mu}H^{-} \right) \, ,
    \label{Eq:Vertex_c3} \\
    \nonumber \\
    H_{1}^{\dagger}H_{2} \left( D_{\mu} H_{2} \right)^{\dagger} D^{\mu}H_{1} + \mathrm{h.c.} &\supset \frac{1}{2}\partial_{\mu}H \left( H h \partial^{\mu}h + v H \partial^{\mu}h  + H G \partial^{\mu}G - v A \partial^{\mu}G \right) \nonumber\\
    &\hspace{5mm} + \frac{v}{2} H \left( \partial_{\mu}G \partial^{\mu}A + \partial_{\mu}G^{-} \partial^{\mu}H^{+} + \partial_{\mu}G^{+} \partial^{\mu}H^{-} \right) \, ,
    \label{Eq:Vertex_c4}
\end{align}
and the corresponding cross sections in Fig.~\ref{Fig:ThermalCrossSection_c34Dependence} exhibit a different behaviour than the previous ones.
The operators presented above are almost identical: The differences appear in interaction terms involving the neutral Goldstone boson~$G$. Hence, the Wilson coefficients $C_{3}$ and $C_{4}$ affect the cross section of $HH\rightarrow ZZ$ in an \emph{asymmetric} way, whereas they influence the cross sections of the other annihilation processes in a \emph{symmetric} way. As a result, the latter cross sections are invariant under an exchange of Wilson coefficients.  While the cross sections for $HH\rightarrow hh$ can easily be understood from the four-point interaction, the particular behaviour of the cross sections of DM annihilations into gauge bosons requires the interplay of $s$-, $t$- and possibly $u$-channels to obtain the cancellation.

\subsection{Impact on the direct-detection cross section $\sigma_{\mathrm{SI}}$}
\begin{figure}[b!]
    \centering
    \includegraphics[width=0.48\textwidth]{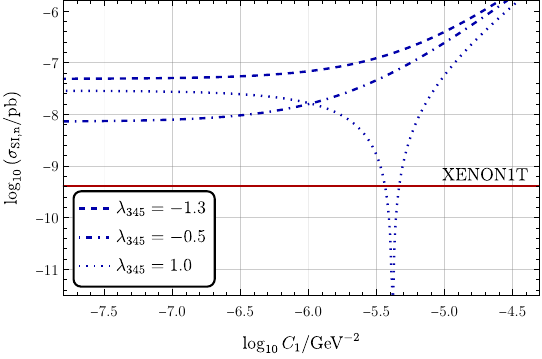}
    \includegraphics[width=0.48\textwidth]{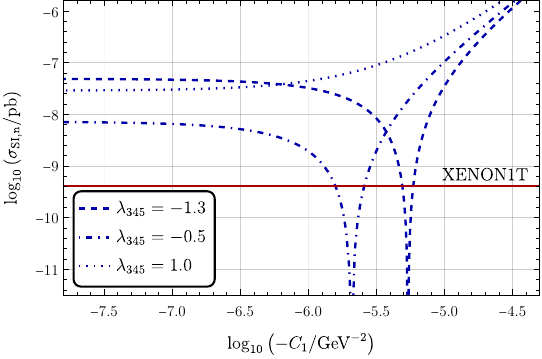}
    \caption{Dependence of the neutron SI DD cross section $\sigma_{\mathrm{SI,n}}$ on the sign of $C_{1}$ for the DM mass $m_{H}=490\GeV$, the mass splitting $\Delta m = 120\GeV$, and the Higgs portal coupling $\lambdaTFF$ as displayed.}
    \label{Fig:SICrossSection_c1Dependence}
\end{figure}

One can ask whether and to what extent these operators affect the spin-independent (SI) direct-detection (DD) cross section~$\sigma_{\mathrm{SI}}$, which are mediated solely by an SM Higgs boson. Besides the vanilla IDM vertex factor and a term proportional to $C_{1}$, we obtain
\begin{align}
    \left( C_{3} + C_{4} \right) \left( p_{1,\mu} p_{2}^{\mu} - p_{1,\mu} p_{3}^{\mu} \right) \propto p_{1,\mu} p_{1}^{\mu}
\end{align}
due to momentum conservation for the contribution to the $HHh$-vertex factor, with $p_{1}$ and $p_{2,3}$ being the SM Higgs' and the DM particles' momenta, respectively. Since \micromegas computes $\sigma_{\mathrm{SI}}$ in the limit of vanishing square of the momentum transfer, \emph{i.e.} $q^{2} \equiv p_{1,\mu} p_{1}^{\mu} = 0$, contributions of the operators associated with $C_{3,4}$ are absent in our numerical results. This approximation is justified as the transferred momentum in DD scattering processes is $\mathcal{O} \left( 1 - 100 \right)\MeV \ll m_{h}$~\cite{Zyla_2020}. Hence, the only contribution to the SI DD cross section arises from the first operator and the dependence of $\sigma_{\mathrm{SI}}$ on the sign of $C_{1}$ and the Higgs portal coupling is shown in Fig.~\ref{Fig:SICrossSection_c1Dependence}.

\clearpage

\section{Remarks on UV-complete models}\label{App:UVcompletion}
\subsection{UV realization in the low-mass regime}

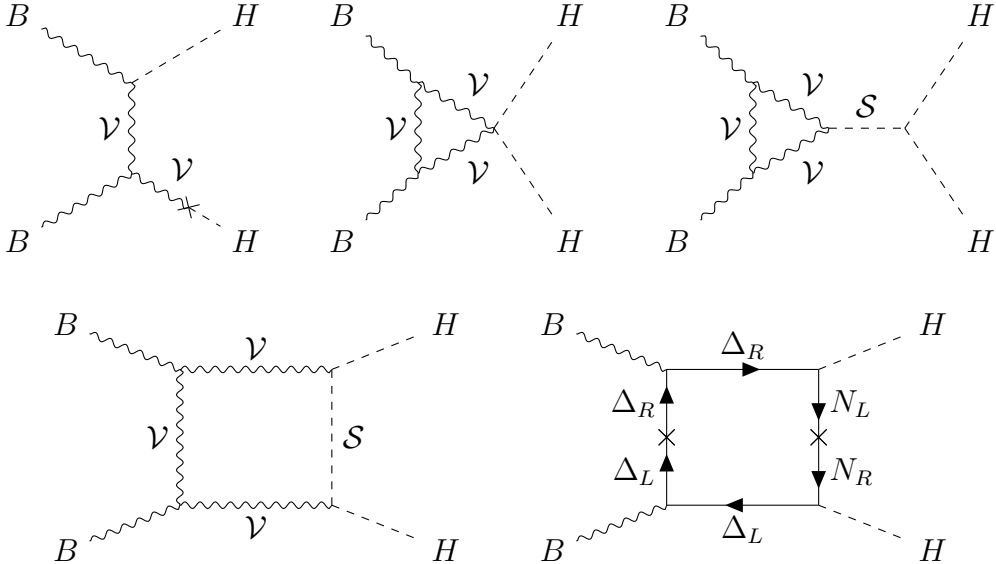
\begin{figure}[b!]
\centering
\begin{tikzpicture}
\begin{feynman}
\vertex (a1) {\(B\)};
\vertex[above=3cm of a1] (a2) {\(B\)};
\vertex at ($(1.5cm,0) + (a1)!0.3!(a2)$) (b1);
\vertex at ($(1.5cm,0) + (a2)!0.3!(a1)$) (b2);

\vertex at ($(a1) + (3cm,0)$) (d1) {\(H\)};
\vertex at ($(a2) + (3cm,0)$) (d2) {\(H\)};

\vertex at ($(b1)!0.5!(d1)$) (c1);

\diagram* {
{[edges=boson]
(a1) -- (b1) -- [edge label=\(\mathcal{V}\)] (b2) -- (a2)
},
(b1) -- [boson, insertion=1, edge label=\(\mathcal{V}\)] (c1),
(c1) -- [scalar] (d1),
(b2) -- [scalar] (d2)
};
\end{feynman}
\end{tikzpicture}\hspace{5mm}
\begin{tikzpicture}
\begin{feynman}
\vertex (a1) {\(B\)};
\vertex[above=3cm of a1] (a2) {\(B\)};
\vertex at ($(1.0cm,0) + (a1)!0.3!(a2)$) (b1);
\vertex at ($(1.0cm,0) + (a2)!0.3!(a1)$) (b2);
\vertex at ($(1.0cm,0) + (b1)!0.5!(b2)$) (c1);

\vertex at ($(a1) + (3cm,0)$) (d1) {\(H\)};
\vertex at ($(a2) + (3cm,0)$) (d2) {\(H\)};

\diagram* {
{[edges=boson]
(a1) -- (b1) -- [edge label=\(\mathcal{V}\)] (b2) -- (a2), (b1) -- [edge label'=\(\mathcal{V}\)] (c1) -- [edge label'=\(\mathcal{V}\)] (b2)
},
(c1) -- [scalar] (d1),
(c1) -- [scalar] (d2)
};
\end{feynman}
\end{tikzpicture}
\hspace{5mm}
\begin{tikzpicture}
\begin{feynman}
\vertex (a1) {\(B\)};
\vertex[above=3cm of a1] (a2) {\(B\)};
\vertex at ($(1.0cm,0) + (a1)!0.3!(a2)$) (b1);
\vertex at ($(1.0cm,0) + (a2)!0.3!(a1)$) (b2);
\vertex at ($(1.0cm,0) + (b1)!0.5!(b2)$) (c1);

\vertex at ($(1cm,0) + (c1)$) (d1);

\vertex at ($(a1) + (4cm,0)$) (e1) {\(H\)};
\vertex at ($(a2) + (4cm,0)$) (e2) {\(H\)};

\diagram* {
{[edges=boson]
(a1) -- (b1) -- [edge label=\(\mathcal{V}\)] (b2) -- (a2), (b1) -- [edge label'=\(\mathcal{V}\)] (c1) -- [edge label'=\(\mathcal{V}\)] (b2)
},
(c1) -- [scalar, edge label=\(\mathcal{S}\)] (d1),
(d1) -- [scalar] (e1),
(d1) -- [scalar] (e2)
};
\end{feynman}
\end{tikzpicture}\\
\vspace{5mm}
\begin{tikzpicture}
\begin{feynman}
\vertex (a1) {\(B\)};
\vertex[above=3cm of a1] (a2) {\(B\)};
\vertex at ($(1.5cm,0) + (a1)!0.2!(a2)$) (b1);
\vertex at ($(1.5cm,0) + (a2)!0.2!(a1)$) (b2);
\vertex at ($(2.0cm,0) + (b1)$) (c1);
\vertex at ($(2.0cm,0) + (b2)$) (c2);
\vertex at ($(5.0cm,0) + (a1)$) (d1) {\(H\)};
\vertex at ($(5.0cm,0) + (a2)$) (d2) {\(H\)};
\diagram* {
{[edges=boson]
(a1) -- (b1) -- [edge label'=\(\mathcal{V}\)] (c1), (a2) -- (b2) -- [edge label=\(\mathcal{V}\)] (c2), (b1) -- [edge label=\(\mathcal{V}\)] (b2)
},
(c1) -- [scalar, edge label'=\(\mathcal{S}\)] (c2),
(c1) -- [scalar] (d1),
(c2) -- [scalar] (d2)
};
\end{feynman}
\end{tikzpicture} \hspace{5mm}
\begin{tikzpicture}
\begin{feynman}
\vertex (a1) {\(B\)};
\vertex[above=3cm of a1] (a2) {\(B\)};
\vertex at ($(1.5cm,0) + (a1)!0.2!(a2)$) (b1);
\vertex at ($(1.5cm,0) + (a2)!0.2!(a1)$) (b3);
\vertex at ($(b1)!0.5!(b2)$) (b2);
\vertex at ($(2.0cm,0) + (b1)$) (c1);
\vertex at ($(2.0cm,0) + (b3)$) (c3);
\vertex at ($(c1)!0.5!(c2)$) (c2);
\vertex at ($(5.0cm,0) + (a1)$) (d1) {\(H\)};
\vertex at ($(5.0cm,0) + (a2)$) (d2) {\(H\)};
\diagram* {
{[edges=anti fermion]
(b1) -- [edge label'=\(\Delta_{L}\)] (c1) -- [edge label'=\(N_{R}\)] (c2) -- [edge label'=\(N_{L}\), insertion=0] (c3) -- [edge label'=\(\Delta_{R}\)] (b3) -- [edge label'=\(\Delta_{R}\)] (b2)-- [edge label'=\(\Delta_{L}\), insertion=0] (b1)
},
(a1) -- [boson] (b1),
(a2) -- [boson] (b3),
(c1) -- [scalar] (d1),
(c3) -- [scalar] (d2)
};
\end{feynman}
\end{tikzpicture}
\caption{Example diagrams for realizing the effective interactions relevant for the low-mass regime, including a vector~$\mathcal{V}_{\mu}$, a scalar singlet~$\mathcal{S}$, or vector-like fermions $N$, $\Delta$.}
\label{Fig:UV_completion_LowMass}
\end{figure}

To realize the effective $CP$-violating operator $H_{2}^{\dagger}H_{2} B_{\mu\nu} \widetilde{B}^{\mu\nu}$ (and similarly with $W^{a}_{\mu\nu}$) in Eq.~\eqref{eq:dual-dualOperator}, crucial for the low-mass regime, there are various possibilities, in particular at the one-loop level, which is sufficient to generate the required magnitude of $\tilde{c}_{2} \sim 25 \PeV^{-2}$, found in Sec.~\ref{Sec:EWBG}. Examples of UV realizations are depicted in Fig.~\ref{Fig:UV_completion_LowMass}.

To generate the operator at tree-level, one can introduce a heavy spin-1 field $\mathcal{V}_{\mu}$ in the $\left( 1, \mathbf{2} \right)_{1}$ representation of the SM (with $Q = T_{3} + Y/2$). On top of this, one can for example add a scalar singlet $\mathcal{S}$, which allows for various loop-generated contributions, or envisage vector-like fermions that also generate the operator at one-loop level (see below). 
We note that the operator in Eq.~\eqref{eq:dual-dualOperator} captures all such UV completions via a single new parameter. Extending the results of Ref.~\cite{deBlas:2017xtg} to the IDM, we find that it can be induced from the following bosonic terms (see Fig.~\ref{Fig:UV_completion_LowMass})
\begin{align}
    -\mathcal{L}_{\mathcal{V}, \mathcal{S}} \supset & \left[ \gamma_{\mathcal{V}}  \mathcal{V}_{\mu}^{\dagger} D^{\mu} H_{2} + \mathrm{h.c.} \right] + i g_{\mathcal{V}}^{W} \mathcal{V}_{\mu}^{\dagger} \sigma^{a} \mathcal{V}_{\nu} W^{a\mu\nu}+ i g_{\mathcal{V}}^{\widetilde{W}} \mathcal{V}_{\mu}^{\dagger} \sigma^{a} \mathcal{V}_{\nu} \widetilde{W}^{a\mu\nu} \nonumber \\ \nonumber
    \\
    &+i g_{\mathcal{V}}^{B} \mathcal{V}_{\mu}^{\dagger} \mathcal{V}_{\nu} B^{\mu\nu}+ i g_{\mathcal{V}}^{\widetilde{B}} \mathcal{V}_{\mu}^{\dagger} \mathcal{V}_{\nu} \widetilde{B}^{\mu\nu} + \varepsilon_{\mathcal{S}\mathcal{V}} \mathcal{S} \mathcal{V}_{\mu}^{\dagger} \mathcal{V}^{\mu} + \kappa_{\mathcal{S}}  \mathcal{S} H_{2}^{\dagger} H_{2} \nonumber \\ \nonumber
    \\
    &+ h_{\mathcal{V}}^{(1)} \mathcal{V}_{\mu}^{\dagger} \mathcal{V}^{\mu} H_{2}^{\dagger} H_{2} + h_{\mathcal{V}}^{(2)} \left( \mathcal{V}_{\mu}^{\dagger} H_{2} \right) \left( H_{2}^{\dagger} \mathcal{V}^{\mu}\right) + \left[ h_{\mathcal{V}}^{(3)} \left( \mathcal{V}_{\mu}^{\dagger} H_{2} \right) \left( \mathcal{V}^{\dagger \mu} H_{2} \right) + \mathrm{h.c.} \right] \nonumber \\ \nonumber
    \\
    &+ \left[ g_{\mathcal{S}\mathcal{V}} H_{2}^{\dagger} \left( D_{\mu} \mathcal{S} \right) \mathcal{V}^{\mu} + g_{\mathcal{S}\mathcal{V}}^{\prime} \left( D_{\mu} H_{2} \right)^{\dagger} \mathcal{S} \mathcal{V}^{\mu} + \mathrm{h.c.} \right] \, .
\end{align}
Note that the heavy vector must transform in the same way under a $\mathbb{Z}_{2}$ transformation as the inert Higgs doublet $H_{2}$ to allow for tree-level generation. In addition, this UV extension naturally gives rise to the $CP$-conserving operator $H_{2}^{\dagger}H_{2} W^{a}_{\mu\nu} W^{a\mu\nu}$ (similarly for $B_{\mu\nu}$) through the first line or via loop-suppressed realizations via the other $CP$-conserving operators. Interestingly, this term receives contributions from an additional diagram, compared to the $CP$-violating one, being proportional to $|\gamma_{\mathcal{V}}|^2$ -- which could motivate its larger size (see the discussion in Sec.~\ref{Sec:DM}).\footnote{This UV completion in principle also generates the operator $H_{2}^\dagger \sigma^{a} H_2 \widetilde W_{\mu\nu}^{a} B^{\mu\nu}$. Here we just assume a coupling structure where its coefficient vanishes. Its inclusion would not change our results qualitatively.}

Another possibility is introducing vector-like fermions with appropriate $\mathbb{Z}_{2}$ and weak hypercharge. As an example, we consider the vector-like fermions $N_{(L,R)}:~\left( 1, \mathbf{1} \right)_{0}$ and $\Delta_{(L,R)}:~\left( 1, \mathbf{2} \right)_{-1}$. The relevant part of the Lagrangian reads
\begin{align}
    -\mathcal{L}_{N, \Delta} \supset -\overline{N} i\slashed{D} N  - \overline{\Delta} i\slashed{D} \Delta + \left[ \lambda_{N \Delta}\, \overline{N}_R \tilde{H}_{2}^{\dagger} \Delta_L + \lambda_{\Delta N}\, \overline{\Delta}_R H_{2} N_L + \mathrm{h.c.} \right] \,
\end{align}
and the diagram inducing the $CP$-violating operator in question is depicted in the lower right panel of Fig.~\ref{Fig:UV_completion_LowMass}. Note that there are further fermionic UV completions, for example the fermionic singlet could also be replaced by a triplet.

\subsection{UV realization in the high-mass regime}
\begin{figure}[h!]
\centering
\begin{tikzpicture}
\begin{feynman}
\vertex (a1) {\(H\)};
\vertex[above=3cm of a1] (a2) {\(H\)};
\vertex at ($(1.5cm,0) + (a1)!0.3!(a2)$) (b1);
\vertex at ($(1.5cm,0) + (a2)!0.3!(a1)$) (b2);

\vertex at ($(a1) + (3cm,0)$) (c1) {\(h\)};
\vertex at ($(a2) + (3cm,0)$) (c2) {\(h\)};

\diagram* {
{[edges=scalar]
(a1) -- (b1) -- (c1), (a2) -- (b2) -- (c2)
},
(b1) -- [boson, edge label'=\(\mathcal{B}/\mathcal{W}^{a}\)] (b2)
};
\end{feynman}
\begin{feynman}
\vertex[right=2cm of c1] (d1) {\(H\)};
\vertex[above=3cm of d1] (d2) {\(H\)};

\vertex at ($(d1) + (4.5cm,0)$) (g1) {\(h\)};
\vertex at ($(d2) + (4.5cm,0)$) (g2) {\(h\)};

\vertex at ($(1.5cm,0) + (d1)!0.5!(d2)$) (e1);
\vertex at ($(-1.5cm,0) + (g1)!0.5!(g2)$) (f1);

\diagram* {
{[edges=scalar]
(d1) -- (e1), (d2) -- (e1), (g1) -- (f1), (g2) -- (f1)
},
(e1) -- [boson, edge label'=\(\mathcal{B}^{\prime}/\mathcal{W}^{\prime a}\)] (f1)
};
\end{feynman}
\end{tikzpicture}
\caption{Example diagrams for realizations of the effective vertices in Eq.~\eqref{DerOp}, induced by a heavy vector singlet~$\mathcal{B}^{\left( \prime \right)}$ or vector triplet~$\mathcal{W}^{\left( \prime \right) a}$.}
\label{Fig:UV_completion_HighMass}
\end{figure}
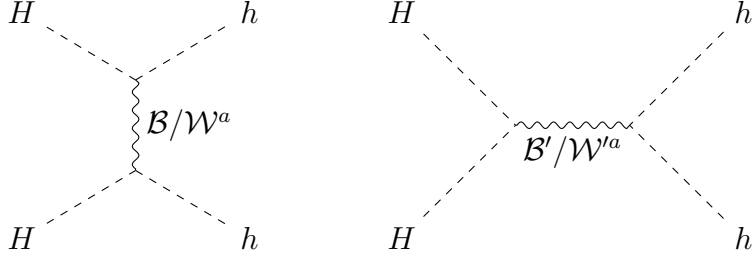

The four effective operators we consider for the high-mass regime are (see 
    Eq.~\eqref{eq:derivativeOperators})
\begin{align}
    \mathcal{L} \supset \ &C_{1}\vert H_{1} \vert^{2} \left( D_{\mu} H_{2} \right)^{\dagger}\left( D^{\mu} H_{2} \right) + C_{2} \vert H_{2} \vert^{2} \left( D_{\mu} H_{1} \right)^{\dagger}\left( D^{\mu} H_{1} \right) \notag\\
    \notag \\
    &+ \left( C_{3} H_{1}^{\dagger}H_{2} \left( D_{\mu} H_{1} \right)^{\dagger}\left( D^{\mu} H_{2} \right) + C_{4} H_{1}^{\dagger}H_{2} \left( D_{\mu} H_{2} \right)^{\dagger}\left( D^{\mu} H_{1} \right) + \mathrm{h.c.} \right) \ .
    \label{DerOp}
\end{align}
Respecting the $\mathbb{Z}_{2}$ symmetry of the inert doublet, two examples for UV realizations are a vector singlet $\mathcal{B}_{\mu}$ and a vector triplet $\mathcal{W}_{\mu}^{a}$ in the representations $\left( 1, \mathbf{1} \right)_{0}$ and $\left( 1, \mathbf{3} \right)_{0}$, respectively. In order to generate the first three operators they must be odd under the same $\mathbb{Z}_{2}$ symmetry as $H_{2}$. The relevant terms read
\begin{align}
    -\mathcal{L}_{\mathcal{B}, \mathcal{W}} \supset &\left[\left( g_{\mathcal{B}}^{D1} \right) \mathcal{B}^{\mu} H_{2}^{\dagger} iD_{\mu} H_{1} + \left( g_{\mathcal{B}}^{D2} \right) \mathcal{B}^{\mu} H_{1}^{\dagger} iD_{\mu} H_{2} + \mathrm{h.c.} \right]\notag\\
   \notag \\
    &+ \left[ \left( g_{\mathcal{W}}^{D1} \right) \mathcal{W}^{a\mu} H_{2}^{\dagger} \frac{\sigma^{a}}{2} iD_{\mu} H_{1} + \left( g_{\mathcal{W}}^{D2} \right) \mathcal{W}^{a\mu} H_{1}^{\dagger} \frac{\sigma^{a}}{2} iD_{\mu} H_{2} + \mathrm{h.c.} \right] \ .
\end{align}
Our fourth operator, though, requires the same set of operators but with new vector fields which are \emph{even} under the $\mathbb{Z}_{2}$ symmetry since either Higgs doublet appears \emph{twice}. The modified set of operators reads
\begin{align}
    -\mathcal{L}_{\mathcal{B}^{\prime}, \mathcal{W}^{\prime}} \supset &\left[\left( g_{\mathcal{B^{\prime}}}^{D1} \right) \mathcal{B^{\prime}}^{\mu} H_{1}^{\dagger} iD_{\mu} H_{1} + \left( g_{\mathcal{B^{\prime}}}^{D2} \right) \mathcal{B^{\prime}}^{\mu} H_{2}^{\dagger} iD_{\mu} H_{2} + \mathrm{h.c.} \right]\notag\\
   \notag \\
    &+ \left[ \left( g_{\mathcal{W^{\prime}}}^{D1} \right) \mathcal{W^{\prime}}^{a\mu} H_{1}^{\dagger} \frac{\sigma^{a}}{2} iD_{\mu} H_{1} + \left( g_{\mathcal{W^{\prime}}}^{D2} \right) \mathcal{W^{\prime}}^{a\mu} H_{2}^{\dagger} \frac{\sigma^{a}}{2} iD_{\mu} H_{2} + \mathrm{h.c.} \right]
\end{align}
and the primed heavy vectors are in the same representations as their siblings above. Relevant diagrams for the matching are depicted in Fig.~\ref{Fig:UV_completion_HighMass}.

\end{appendix}

\clearpage

\bibliographystyle{JHEP} 
\bibliography{Extended_IDM_Bib}

\end{document}